\newcommand\vx{\vec{x}}
\newcommand\vr{\vec{r}}
\newcommand\vk{\vec{k}}

\newcommand\hk{\hat{k}}

\newcommand\hz{\hat{z}}

\newcommand\hr{\hat{r}}

\newcommand\oO{\mathcal{O}}

\newcommand\Mpc{\rm\; Mpc}
\newcommand\Mpch{{\rm\; Mpc}/h}

\newcommand{\mD}{\mathcal{D}}
\newcommand{\tF}{\tilde{F}_2}
\newcommand{\tG}{\tilde{G}_2}
\newcommand{\mG}{\mathcal{G}(k_1,k_2)}
\newcommand{\vbc}{v_{\rm bc}}

\newcommand{\tdelta}{\tilde{\delta}}

\documentclass[useAMS,usenatbib]{mn2e}

\usepackage{scalerel,stackengine}
\stackMath
\newcommand\reallywidehat[1]{%
\savestack{\tmpbox}{\stretchto{%
  \scaleto{%
    \scalerel*[\widthof{\ensuremath{#1}}]{\kern-.6pt\bigwedge\kern-.6pt}%
    {\rule[-\textheight/2]{1ex}{\textheight}}
  }{\textheight}%
}{0.5ex}}%
\stackon[1pt]{#1}{\tmpbox}%
}

\usepackage{lineno}

\usepackage{placeins}

\usepackage{graphicx} 
\usepackage{multirow}
\usepackage{array}

\usepackage{amsmath,amssymb}

\usepackage[T1]{fontenc}
\usepackage[latin9]{inputenc}
\usepackage{float}
\usepackage{graphicx}
\usepackage{esint}

\makeatletter
\DeclareFontEncoding{LGR}{}{}
\DeclareTextSymbol{\~}{LGR}{126}

\begin{document}

\title[Redshift-space 3PCF]{Modeling the large-scale redshift-space 3-point correlation function of galaxies}

\author{\makeatauthor}

\author[Slepian \& Eisenstein]{Zachary Slepian$^{1}$\thanks{zslepian@cfa.harvard.edu} \& Daniel J. Eisenstein$^{1}$\thanks{deisenstein@cfa.harvard.edu}}

\maketitle

\begin{abstract}
We present a configuration-space model of the large-scale galaxy 3-point correlation function (3PCF) based on leading-order perturbation theory and including redshift space distortions (RSD).  This model should be useful in extracting distance-scale information from the 3PCF via the Baryon Acoustic Oscillation (BAO) method. We include the first redshift-space treatment of biasing by the baryon-dark matter relative velocity. Overall, on large scales the effect of RSD is primarily a renormalization of the 3PCF that is roughly independent of both physical scale and triangle opening angle; for our adopted $\Omega_{\rm m}$ and bias values, the rescaling is a factor of $\sim 1.8$. We also present an efficient scheme for computing 3PCF predictions from our model, important for allowing fast exploration of the space of cosmological parameters in future analyses.

\end{abstract}

\section{Introduction}
The clustering of pairs of galaxies, quantified by the 2-point correlation function (2PCF), has a distinctive peak at roughly $100\Mpch$, an imprint of the Baryon Acoustic Oscillations (BAO) occurring prior to recombination (Sakharov 1966; Peebles \& Yu 1970; Sunyaev \& Zel'dovich 1970; Bond \& Efstathiou 1984, 1987; Holtzmann 1989; Hu \& Sugiyama 1996; Eisenstein \& Hu 1998; Eisenstein, Seo \& White 2007).  The acoustic peak has been successfully used as a standard ruler to measure the relative size of the Universe at different redshifts as well as the absolute expansion rate (by comparison to the Cosmic Microwave Background), and these measurements in turn constrain the cosmological parameters, particularly dark energy (Eisenstein, Hu \& Tegmark 1998b; Blake \& Glazebrook 2003; Hu \& Haiman 2003; Linder 2003; Seo \& Eisenstein 2003; Weinberg et al. 2012, for a review; Cuesta et al. 2015 for the most recent measurement).  

Theoretical modeling indicates that the 3-point correlation function (3PCF) of galaxies should also contain BAO features (Sefusatti et al. 2006; Gil-Mar\'in et al. 2012; Slepian \& Eisenstein 2015a, Slepian et al. 2016a; hereafter SE15a and S16a), and indeed hints have been detected in Gazta\~naga et al. (2009) and S16a.  If detected at high significance, BAO features in the 3PCF could be used as a standard ruler exactly as is already done with the 2PCF. Given the large spectroscopic datasets presently extant, such as BOSS (Eisenstein et al. 2011; Dawson et al. 2013), 6dFGS (Jones et al. 2009), and WiggleZ (Drinkwater et al. 2010), as well as the even-larger ones planned for the coming decade, such as those from eBOSS,  Dark Energy Spectroscopic Instrument (DESI; Levi et al. 2013) and Wide-Field Infrared Survey Telescope (WFIRST; Spergel et al. 2013), it will be desirable to exploit the 3PCF as an additional tool for constraining the cosmic expansion history.

However, the true positions of galaxies along our line-of-sight are unknown, and the redshift is used as a proxy, converted to a distance by assuming that the galaxies are comoving with the background Universe's expansion and have no peculiar velocities. This assumption is not perfectly correct. On small scales, galaxies have thermal peculiar velocities due to virialization within clusters. On larger scales, the peculiar velocities are coherent and generated by the growth of structure.  Consequently, the positions of galaxies inferred from redshifts will be subject to redshift-space distortions (RSD; Hamilton 1998, for a review) unless these peculiar velocities are accounted for. In the 2PCF, RSD are modeled using the Kaiser formula (Kaiser 1987; Hamilton 1992), which rescales the 2PCF monopole by $1+2\beta/3 + \beta^2/5$, where $\beta = f/b_1$, with $f\approx \Omega_{\rm m}^{0.55}$ the logarithmic derivative of the linear growth rate with respect to scale factor and $b_1$ the linear bias. The Kaiser formula assumes a constant, single line of sight to the entire survey, which appears to be sufficiently accurate for the Sloan Digital Sky Survey (SDSS) volume (see Slepian \& Eisenstein 2015e and references therein for further discussion). Because the Kaiser factor is scale-independent, it does not shift the BAO scale in the 2PCF used to measure the expansion rate.  The small-scale thermal velocities additionally alter the 2PCF but can be modeled as a Gaussian smoothing of the power spectrum, which also does not shift the BAO scale.  

For the 3PCF, there has as yet been no configuration space model from perturbation theory.   Scoccimarro, Frieman \& Couchman (1999; hereafter SCF99) use Standard Perturbation Theory (SPT) working to second-order in the linear density field $\delta$ to obtain a model of the bispectrum (Fourier analog of the 3PCF). Rampf \& Wong (2012) show that the tree-level Lagrangian perturbation theory result, from using the Zel'dovich approximation, agrees with the SPT result of SCF99.  There have been a number of models for the redshift-space bispectrum, both fully analytic (e.g. Hivon et al. 1995; Verde, Heavens \& Matarrese 1998; Smith, Sheth \& Scoccimarro 2008) and incorporating numerical simulations (e.g. Gil-Mar\'in et al. 2014). However only very few works consider the BAO. Sefusatti et al. (2006) focuses on joint analysis of the power spectrum and bispectrum and notes that BAO can break degeneracies (see their Figures 7, 9, and 10). Gil-Mar\'in et al. (2012) give a fitting formula for the dark matter bispectrum including BAO, and Gil-Mar\'in et al. (2014) includes RSD.

The purpose of the present work is to develop a model of the 3PCF in configuration space in a form suitable for fitting the 3PCF of a  large-scale redshift survey. First, we will convert the bispectrum model of SCF99 to configuration space. We will find that RSD essentially rescale the no-RSD 3PCF in a way that is roughly independent of both physical scale and triangle opening angle.  This conclusion develops ideas first advanced in S16a and helps explain why the configuration-space model without RSD in that work was able to obtain a reasonable fit to the data.  In the present work, we also develop a redshift-space model of the baryon-dark matter relative velocity effect, developing previous work on this term's signature in the 3PCF in real-space (SE15a).

As a second goal of this paper, we will present a fast scheme for computing 3PCF predictions in the multipole basis first proposed in Szapudi (2004) and further developed in Slepian \& Eisenstein (2015a, b, c; hereafter SE15a, b, c). Typically perturbation theory expressions for the 3PCF $\zeta$ are written as cyclic sums over functions of pairs of sides and their enclosed angle, for instance in the form $\zeta\sim \xi(r_1)\xi(r_2) +{\rm cyc.}$, with $\xi$ the 2PCF (Groth \& Peebles' 1977 ``hierarchical ansatz''; see also Fry \& Peebles 1978; Davis \& Peebles 1977; Ma \& Fry 2000).  Each term in the cyclic sum of such an expression corresponds to a different galaxy's contributing a particular bias term to the expectation value $\left< \delta_{\rm g}\delta_{\rm g}\delta_{\rm g}\right>$, as we further explain in \S\ref{sec:cyc_sum}.

In reality, it is unknown which galaxy contributes which bias term, and one must cyclically sum so that all galaxies have a chance to contribute all the bias terms relevant at a given order in perturbation theory.  Given two sides $r_1$ and $r_2$ and the cosine of their enclosed angle, $\hr_1\cdot \hr_2$, cyclic summing requires computing the third side and the two additional angles.  This side and these angles depend on non-separable functions of $r_1,\;r_2$, and $\hr_1\cdot\hr_2$ and so their calculation scales as the number of grid points used for each side, $N_r$, times the number of grid points in angle cosine, $N_{\mu}$---that is, $N_r^2N_{\mu}$.

Yet in the end we wish to bin the predictions in side lengths to a relatively modest number of bins, $N_{\rm bins}$, and also project the angular dependence onto Legendre polynomials. In this work we show how to do these operations first, meaning that the cyclic summing can be made to scale as $N_{\rm bins}^2$ for each multipole, for a total scaling as $N_{\rm bins}^2\ell_{\rm max}$ with $\ell_{\rm max}$ the maximal multipole. Computing the 3PCF predictions in the multipole basis using this scheme is consequently significantly more efficient. This efficiency will be important as the 3PCF becomes a standard tool for large-scale structure analyses and it becomes desirable to run a large grid of cosmological parameters through a prediction pipeline.

The paper is laid out as follows. In \S\ref{sec:RSD_model}, we present the redshift-space bispectrum model of SCF99 and show how to cast it to configuration space. We then incorporate add tidal tensor biasing and briefly discuss other possible refinements to our model. In \S\ref{sec:cyc_sum}, we present the more efficient cyclic summing scheme summarized above.  \S\ref{sec:disc} discusses our results after cyclic summing, and \S\ref{sec:rv_contrib} shows how to add relative velocity biasing in redshift space. \S\ref{sec:concs} concludes. Two Appendices showing mathematical results used in the main text follow \S\ref{sec:concs}. 

For all of the results displayed in this work, we have used transfer functions output from \textsc{CAMB} (Lewis 2000) with a geometrically flat $\Lambda{\rm CDM}$ cosmology with the following parameters: $\Omega_{\rm b}h^2 = 0.0220453,\;\Omega_{\rm c}h^2 = 0.119006,\;T_{\rm CMB} = 2.7255\;{\rm K}$, $n_{\rm s} = 0.9611$. These parameters match those used in S16a and do not differ substantially from the Planck values (Planck Paper XIII, 2015).  Our $\sigma_8(z=0) = 0.8288$, and we rescale $\sigma_8$ by the ratio of the linear growth factor at the survey redshift to the linear growth factor at redshift zero. We take the survey redshift to be $z_{\rm survey} = 0.565$ so that our results are comparable to the CMASS galaxy sample discussed in S16a.

\section{RSD model with linear and non-linear biasing}
\label{sec:RSD_model}
\subsection{Multipoles in Fourier space: pre-cyclic}

For an idealized survey (constant line of sight to the survey, uniform density, etc.), the full redshift-space bispectrum depends on five parameters: three to characterize the triangle's shape, e.g. two sides and the enclosed angle, and two to describe the orientation of the triangle to the line of sight.  One starts with nine parameters describing each coordinate of the three triangle vertices; translation invariance reduces this to six and rotation invariance about the line of sight to five.  

SCF99 uses the angle of one triangle side (in Fourier space) to the line of sight and the azimuthal angle of the second side about this first side to capture the orientation. SCF99 averages over all azimuthal angles of the second side about the first side to write the redshift-space bispectrum as a multipole series with angular piece dependent on the angle between the line of sight and the first side.  Further averaging over all orientations of this first side selects the monopole moment in their equation (20). Our focus here is the fully averaged 3PCF, so this monopole moment is our starting point. It is
\begin{align}
B_{\rm s}(k_1,k_2,x)=&b_1^3P(k_1)P(k_2)\bigg[\tF(k_1,k_2;x)\mD_{\rm SQ1}(\beta,x) +\nonumber\\ &\tG(k_1,k_2;x)\mD_{\rm SQ2}(\beta,k_1,k_2;x)+\nonumber\\
&\mD_{\rm NLB}(\beta,\gamma;x)+\mD_{\rm FOG}(\beta,k_1,k_2;x)\bigg]+{\rm cyc.},
\label{eqn:model}
\end{align}
where $P$ is the linear theory matter power spectrum, $b_1$ is the linear bias, $\gamma = 2b_2/b_1$ is the ratio\footnote{Our galaxy bias model has a term proportional to $b_2$, while SCF99's uses $b_2/2$, leading to a factor of $2$ in $\gamma$ relative to its definition in SCF99.} of non-linear bias $b_2$ to linear bias, $\beta = f/b_1$, with $f=d\ln D/d\ln a\approx \Omega_{\rm m}^{0.55}$ the logarithmic derivative of the linear growth rate $D$ with respect to scale factor $a$, and $x\equiv \hk_1\cdot\hk_2$. Our notation mostly follows SCF99's; subscript $s$ denotes redshift space; $\tF$ is the second-order density kernel and $\tG$ the second-order velocity kernel, while subscript $SQ1$ denotes linear squashing, $SQ2$ non-linear squashing, $NLB$ non-linear bias, and $FOG$ fingers-of-God. The second-order density and velocity kernels are SCF99 equations (22) and (23), while the various $\mD$ are SCF99 equations (24), (26), (28), and (30).  Since we only consider the monopole with respect to the line of sight here, we have suppressed SCF99's superscript zero in the bispectrum as well as in the $\mD$'s.

As discussed in \S1, for comparison with observational work we seek the configuration-space multipoles of the 3PCF. Here we focus on the pre-cyclic multipoles, deferring cyclic summing around the triangle to \S\ref{sec:cyc_sum}. Ideally, we would write the bispectrum in a multipole series with radial coefficients in $k_1,k_2$ times angular dependences in $x$, i.e.
\begin{align}
B_{\rm s}(k_1,k_2;x)=\sum_{\ell}B_{\rm s,\ell}(k_1,k_2)P_{\ell}(x),
\label{eqn:ideal_multipoles}
\end{align}
where $P_{\ell}$ is a Legendre polynomial. This form is desirable because using results in the Appendix of SE15a one can show that the inverse FT of equation (\ref{eqn:ideal_multipoles}) is
\begin{align}
\zeta_{\rm s}(r_1,r_2;\hr_1\cdot\hr_2)=\sum_{\ell}\zeta_{\ell}(r_1,r_2)P_{\ell}(\hr_1\cdot\hr_2)
\label{eqn:simple_inv_FT}
\end{align}
with 
\begin{align}
&\zeta_{\ell}(r_1,r_2)=\nonumber\\
&(-1)^{\ell}\int\frac{k_1^2k_2^2dk_1dk_2}{(2\pi^2)^2}B_{{\rm s},\ell}(k_1,k_2)j_{\ell}(k_1r_1)j_{\ell}(k_2r_2),
\label{eqn:2d_transform}
\end{align}
which further simplifies to the product of two 1-D integral transforms if the bispectrum's coefficients $B_{{\rm s}, \ell}(k_1,k_2)$ are separable.

Many of the terms in the model of SCF99 can be written in this form. However, some terms in this model involve explicit dependence on $1/k_3^2$, with $k_3=|\vk_1+\vk_2|$.  While one could write a Legendre expansion of $1/k_3^2$ by using the expansion for $1/k_3$ and squaring, this would involve the square of an infinite Legendre series with coefficients of the form $(k_1/k_2)^{\ell}$ for $k_1<k_2$ and the inverse where $k_2<k_1$. Casting such a series to configuration space would require integrals of arbitrary powers of $k_1$ and $k_2$. Consequently if performed separately these integrals will not in general converge, though if performed as a 2-D integral involving the ratio $k_1/k_2$ they will. For both simplicity and computational efficiency, it is desirable to cast the inverse FT of the terms involving $k_3$ as a product of separable 1-D integrals.  Here we will develop a technique to achieve this goal. Details will follow; for the moment having this approach in mind motivates us to write the bispectrum as
\begin{align}
&B_{\rm s}(k_1,k_2;x)=\nonumber\\
&\sum_{\ell}\bigg[B_{{\rm s},\ell,{\rm no}\;k_3}(k_1,k_2)+B_{{\rm s},\ell, {\rm with}\;k_3}(k_1,k_2)\bigg]P_{\ell}(x).
\label{eqn:bispec_decomp_nok3_k3}
\end{align}
The inverse FT of the first term above is straightforwardly given by equations (\ref{eqn:simple_inv_FT}) and (\ref{eqn:2d_transform}), while the second  will require the Delta-function technique outlined above.

We now expand the ``no $k_3$'' terms that result from expanding out equation (\ref{eqn:model}) using the explicit forms for the kernels entering it. We do not reproduce here these kernels because they are lengthy. We focus on the non-equilateral results of SCF99 (their \S3.1) as equilateral triangles are a particular continuous limit of these expressions. To make the calculation easier to connect to SCF99, we address each term in the sum equation (\ref{eqn:model}) separately. The pre-factor $b_1^3 P(k_1)P(k_2)$ in equation (\ref{eqn:model}) is already in a separable form, so we focus on manipulating the terms in the square brackets in equation (\ref{eqn:model}), where again the explicit forms are given in SCF99.

For $\tF(k_1,k_2;x)\mD_{\rm SQ1}(\beta,x)$ we find:
\begin{align}
&\ell =0:\;\frac{34}{21}\left[1+\frac{2}{3}\beta+\frac{49}{425}\beta^2\right]\nonumber\\
&\ell=1:\;\mG \left[1+\frac{2}{3}\beta + \frac{11}{75}\beta^2 \right]\nonumber\\
&\ell=2:\;\frac{8}{21}\left[1+\frac{2}{3}\beta + \frac{18}{35}\beta^2 \right]\nonumber\\
&\ell=3:\;\mG\frac{4}{75}\beta^2\nonumber\\
&\ell=4:\;\frac{64}{3675}\beta^2,
\label{eqn:F2_Dsq1}
\end{align}
where 
\begin{align}
\mG\equiv\left( \frac{k_1}{k_2}+\frac{k_2}{k_1}\right),
\end{align}
with $\mG$ for ``gradient'', as this $k$-dependence is generated by gradients entering the PT, as discussed in Bernardeau et al. (2002), SE15b, and S16a. The no-RSD limit as $\beta\to 0$ recovers the coefficients in SE15a equations (52)-(54) for the pre-cyclic 3PCF (ignoring relative velocity bias $b_v$). Both $\tF$ and $\mD_{\rm SQ1}$ had terms up $x^2$ ($\ell=2$ each), so by angular momentum addition the coefficients reach $\ell=4$.

For $\mD_{\rm NLB}(\beta,\gamma,x)$ we find
\begin{align}
&\ell=0:\;\gamma \left[1+\frac{2}{3}\beta+\frac{1}{9}\beta^2\right]\nonumber\\
&\ell=2:\;\gamma \frac{4}{45}\beta^2
\label{eqn:nlb}
\end{align}
The no-RSD limit as $\beta\to 0$ recovers the correct $b_2$ coefficient in SE15a equation (52). $\mD_{\rm NLB}$ only had constant and $x^2$ terms and so we obtain only $\ell=0$ and $2$ moments.

For $\mD_{\rm FOG}(\beta,k_1,k_2;x)$ we find
\begin{align}
&\ell=0:\;\frac{2}{3}\beta+\frac{38}{45}\beta^2+\frac{2}{5}\beta^3+\frac{2}{25}\beta^4\nonumber\\
&\ell=1:\;\mG\left[\frac{1}{3}\beta+\frac{3}{5}\beta^2+\frac{67}{175}\beta^3+\frac{3}{35}\beta^4\right]\nonumber\\
&\ell=2:\frac{16}{45}\beta^2+\frac{16}{35}\beta^3+\frac{32}{245}\beta^4\nonumber\\
&\ell=3:\;\mG\left[\frac{8}{175}\beta^3+\frac{8}{315}\beta^4\right]\nonumber\\
&\ell=4:\;\frac{128}{11025}\beta^4
\label{eqn:fog}
\end{align}
As $\beta\to0$ there is no contribution from fingers-of-God. $\mD_{\rm FOG}$ had terms up to $x^4$ so the multipole moments reach $\ell=4$. Finally, these fingers-of-God are not from small-scale thermal velocities due to virialization, but are rather large-scale, linear-theory effects; we term them fingers-of-God to follow the usage of SCF99.

We now address the term in equation (\ref{eqn:bispec_decomp_nok3_k3}) involving $k_3$-dependence, which enters solely through $\mD_{\rm SQ2}$. First, we write out $\mD_{\rm SQ2}$ as
\begin{align}
&\mD_{\rm SQ2}(\beta,k_1,k_2;x)=\frac{2}{105}\bigg[\{35\beta+[28+3\beta\nonumber\\
&+2x^2(7+6\beta)]\beta^2\}-\frac{\beta^2}{k_3^2}\left\{4k_1k_2x(x^2-1)(7+3\beta)\right\}\bigg].
\label{eqn:dsq2}
\end{align}
This decomposition is not unique; we can always replace $k_1k_2=(k_3^2-k_1^2-k_2^2)/2$ if we wish. The decomposition above is favorable for keeping the algebra to follow compact.

We now multiply $\mD_{\rm SQ2}$ as written above by $\tG(k_1,k_2;x)$. The result of this multiplication for the terms in the first curly bracket of equation (\ref{eqn:dsq2}), which have no $k_3$-dependence, can then be projected onto multipoles, yielding coefficients
\begin{align}
&\ell=0:\;\frac{26}{63}\beta+\frac{628}{1575}\beta^2+\frac{346}{3675}\beta^3\nonumber\\
&\ell=1:\;\mG\left[\frac{1}{3}\beta+\frac{26}{75}\beta^2+\frac{17}{175}\beta^3\right]\nonumber\\
&\ell=2:\;\frac{16}{63}\beta+\frac{808}{2205}\beta^2+\frac{832}{5145}\beta^3\nonumber\\
&\ell=3:\;\mG\left[\frac{4}{75}\beta^2+\frac{8}{175}\beta^3\right]\nonumber\\
&\ell=4:\;\frac{128}{3675}\beta^2+\frac{256}{8575}\beta^3.
\label{eqn:k3_indpdt}
\end{align}

For the result of multiplying $\tG(k_1,k_2;x)$ by terms in the second curly bracket of equation (\ref{eqn:dsq2}), which have a pre-factor of $1/k_3^2$, we require a strategy for taking the inverse FT. As earlier discussed, while $k_3$ is fully determined by $k_1,k_2$ and $x$, writing it out explicitly in this basis would generate the square of an infinite Legendre series with some associated disadvantages. Consequently we will here instead treat $\vk_3$ as an additional degree of freedom and enforce the constraint that it forms a closed triangle with $\vk_1$ and $\vk_2$ by integrating over $d^3\vk_3$ against a Dirac delta function $\delta^{[3]}_{\rm D}(\vk_1+\vk_2-\vk_3)$. Thus for now we treat $k_3$ as independent of $k_1,k_2$ and $x$. Projecting the result of multiplying $\tG(k_1,k_2;x)$ by terms in the second curly bracket in equation (\ref{eqn:dsq2}) onto Legendre polynomials, we find
\begin{align}
&\ell=0:\;\frac{8}{1575k_3^2}\left(k_1^2+k_2^2\right)\left(7\beta^2+3\beta^3\right)\nonumber\\
&\ell=1:\;\frac{176}{8575k_3^2}\left(k_1k_2\right)\left(7\beta^2+3\beta^3\right)\nonumber\\
&\ell=2:\;\frac{8}{2205k_3^2}\left(k_1^2+k_2^2\right)\left(7\beta^2+3\beta^3\right)\nonumber\\
&\ell=3:\;-\frac{496}{33075k_3^2}\left(k_1k_2\right)\left(7\beta^2+3\beta^3\right)\nonumber\\
&\ell=4:\;-\frac{32}{3675k_3^2}\left(k_1^2+k_2^2\right)\left(7\beta^2+3\beta^3\right)\nonumber\\
&\ell=5:\;-\frac{256}{46305k_3^2}\left(k_1k_2\right)\left(7\beta^2+3\beta^3\right).
\end{align}
We note the alternation of the scale-dependence from even to odd $\ell$.  

Taking the inverse FT of these coefficients would be simplified were all the $k$-dependence reducible to the form $k_1k_2$.  To achieve this reduction, in all of the even-$\ell$ coefficients we replace $k_1^2+k_2^2=k_3^2-2k_1k_2x$. Making this replacement reintroduces $x$-dependence so we must now reproject onto the Legendre polynomials.  This reprojection requires writing out the full set of (even) terms above including their angular dependence.  Because the replacement we make involves $x$, it will couple a given multipole $\ell$ to $\ell\pm1$ upon reprojection. Making the replacement and reprojecting we find the new multipole coefficients from the $k_3$-dependent piece of $\tG\mD_{\rm SQ2}$ are
\begin{align}
&\ell=0:\;\frac{8}{1575}\left(7\beta^2+3\beta^3\right)\nonumber\\
&\ell=1:\; -\frac{16}{1225}\frac{k_1k_2}{k_3^2}\left(7\beta^2+3\beta^3\right)\nonumber\\
&\ell=2:\;\frac{8}{2205}\left(7\beta^2+3\beta^3\right)\nonumber\\
&\ell=3:\;\frac{16}{4725}\frac{k_1k_2}{k_3^2}\left(7\beta^2+3\beta^3\right)\nonumber\\
&\ell=4:\;-\frac{32}{3675}\left(7\beta^2+3\beta^3\right)\nonumber\\
&\ell=5:\;\frac{64}{6615}\frac{k_1k_2}{k_3^2}\left(7\beta^2+3\beta^3\right).
\label{eqn:k3_dpdt}
\end{align}

Finally, we co-add all of the terms together at each multipole, save for those that involve $k_3$. Explicitly, this is the sum of equations (\ref{eqn:F2_Dsq1}),  (\ref{eqn:nlb}), (\ref{eqn:fog}), (\ref{eqn:k3_indpdt}), and the even multipoles only of equation (\ref{eqn:k3_dpdt}). We also now incoporate the factor of $b_1^3 P(k_1)P(k_2)$ of equation (\ref{eqn:model}). We find
\begin{align}
&\ell=0:\;b_1^3 P(k_1)P(k_2)\bigg\{\frac{34}{21}\left[1+\frac{4}{3}\beta+\frac{1154}{1275}\beta^2+\frac{936}{2975}\beta^3+\frac{21}{425}\beta^4\right] \nonumber\\
&\;\;\;\;\;\;\;\;\;\;\;\; + \gamma \left[1+\frac{2}{3}\beta+\frac{1}{9}\beta^2\right]\bigg\}\nonumber\\
&\ell=1:\; b_1^3 P(k_1)P(k_2)\mG\left[1+\frac{4}{3}\beta+\frac{82}{75}\beta^2+\frac{12}{25}\beta^3+\frac{3}{35}\beta^4\right]\nonumber\\
&\ell=2:\;b_1^3 P(k_1)P(k_2)\bigg\{\frac{8}{21}\left[1+\frac{4}{3}\beta+\frac{52}{21}\beta^2+\frac{81}{49}\beta^3+\frac{12}{35}\beta^4\right] \nonumber\\
&\;\;\;\;\;\;\;\;\;\;\;\; +\frac{32\gamma}{945}\beta^2\bigg\}\nonumber\\
&\ell=3:\;b_1^3 P(k_1)P(k_2)\mG\left[\frac{8}{75}\beta^2+\frac{16}{175}\beta^3+\frac{8}{315}\beta^4\right]\nonumber\\
&\ell=4:\;b_1^3 P(k_1)P(k_2)\left[-\frac{32}{3675}\beta^2+\frac{32}{8575}\beta^3+\frac{128}{11025}\beta^4\right].\nonumber\\
\label{eqn:coadded_nok3_multis}
\end{align}
Again, this result does not include those terms in equation (\ref{eqn:k3_dpdt}) with $k_3$-dependence. While in Fourier space the $k_3$-dependent terms enter only the odd-$\ell$ terms in the decomposition we have adopted, these terms will enter both even and odd configuration-space multipoles, as we next show. Incorporating $b_1^3$ and the power spectrum, the $k_3$-dependent contributions are
\begin{align}
&\ell=1:\; -\frac{16}{1225}\left(7\beta^2+3\beta^3\right)b_1^3 P(k_1)P(k_2)\frac{k_1k_2}{k_3^2}\nonumber\\
&\ell=3:\;\frac{16}{4725}\left(7\beta^2+3\beta^3\right)b_1^3 P(k_1)P(k_2)\frac{k_1k_2}{k_3^2}\nonumber\\
&\ell=5:\;\frac{64}{6615}\left(7\beta^2+3\beta^3\right)b_1^3 P(k_1)P(k_2)\frac{k_1k_2}{k_3^2}.
\label{eqn:k3_dpdt_with_powerspec}
\end{align}

The sum of equations (\ref{eqn:coadded_nok3_multis}) and (\ref{eqn:k3_dpdt_with_powerspec}) represents the full decomposition of the line-of-sight averaged SCF99 model into the Legendre basis.

\subsection{Multipoles in configuration space: pre-cyclic}
We now wish to take the inverse FT of the coefficients presented in equations (\ref{eqn:coadded_nok3_multis}) and )(\ref{eqn:k3_dpdt_with_powerspec}). We first focus on equation (\ref{eqn:k3_dpdt_with_powerspec}) for the $k_3$-dependent pieces of the odd multipoles; these will contribute to $all$ multipoles in configuration space and thus are required before writing down any final multipole coefficients, even or odd.

As we show in Appendix A, the contribution of these odd multipoles of equation (\ref{eqn:k3_dpdt}) (having now included the factor of $b_1^3P(k_1)P(k_2)$) to a given multipole $\ell$ in cofiguration space is
\begin{align}
0\leq \ell<\infty :&\;b_1^3\left(7\beta^2+3\beta^3 \right)\kappa_{\ell}(r_1,r_2),
\label{eqn:SQ2_k3dpdt_config}
\end{align}
with
\begin{align}
\kappa_{\ell}(r_1,r_2) & =\frac{64}{77175}
\bigg[9 I_{1\ell}(r_1,r_2) - 14 I_{3\ell}(r_1,r_2)\nonumber\\
&+5 I_{5\ell}(r_1,r_2)\bigg],
\label{eqn:kappa_ell}
\end{align}
\begin{align}
I_{{\mathcal L}\ell}(r_1,r_2) = &\sum_{l_1}(-1)^{l_1+\ell}(2l_1+1)(2\ell+1)\left(\begin{array}{ccc}
l_{1} & \ell & {\mathcal L}\\
0 & 0 & 0
\end{array}\right)^2\nonumber\\
&\times \int r dr f_{\ell l_1}(r_1;r)f_{\ell l_1}(r_2;r),
\label{eqn:IL_result}
\end{align}
and
\begin{align}
f_{\ell l_1}(r_i;r)=\int \frac{k^2dk}{2\pi^2}j_{\ell}(kr_i)j_{l_1}(kr)kP(k).
\label{eqn:ftens_defn}
\end{align}
Relative to Appendix A we have specialized to the case where $u_{\mathcal L}(k_1)=k_1$ and $u_{\mathcal L}(k_2)=k_2$ for $\mathcal{L}=1,3$, and $5$, and zero otherwise; from equation (\ref{eqn:k3_dpdt_with_powerspec}) we see this is the only case here required. We have suppressed the superscript $\mathcal{L}$ on the $f$-tensor for this reason.  

The numerical pre-factor in equation (\ref{eqn:kappa_ell}) is small compared to unity, $64/77175\approx 8.29\times 10^{-4}$. This smallness will mean the $\kappa_{\ell}$ do not contribute strongly to the pre-cyclic 3PCF. Further, as the pre-cyclic multipole $\ell \gg \mathcal{L}$, the 3j-symbols mean that $l_1\sim \ell$, leading to a squeezed triangle in angular momentum space.  In this limit the 3j-symbols for the three different values of $\mathcal{L}$ take on very similar values, and $f_{\ell l_1}$ is independent of $\mathcal{L}$.  Consequently $I_{1\ell} \sim I_{3\ell} \sim I_{5\ell}$. Were the $I_{\mathcal{L} \ell}$ exactly equal, their numerical pre-factors in equation (\ref{eqn:kappa_ell}) are such that $\kappa_{\ell}(r_1, r_2)\to 0$.  Thus for large $\ell \gg \mathcal{L}$, we expect that the pre-cyclic $\kappa_{\ell}(r_1, r_2)$ become small.  This point is important because the series of pre-cyclic $\kappa_{\ell}(r_1, r_2)$ is formally infinite: all $\ell$ values are required. Thus, to compute the post-cyclic $\kappa_{\ell}$, as we will do in \S{\ref{sec:cyc_sum}, one would need to form this infinite series. However, the fact that for large $\ell$ the pre-cyclic terms become small means that in practice it is accurate to truncate the pre-cyclic series at modest $\ell \sim 6$.  To be conservative we use $\ell$ up to $12$, as we will discuss further in \S{\ref{sec:cyc_sum}.

The fact that terms in $k_3$ couple to all multipoles has a simple geometric interpretation. Any characterization of $k_3$ in terms of the other two sides and their enclosed angle will require all $\ell$, because $k_3 = \sqrt{k_1^2+k_2^2 +2k_1k_2x}$, which as earlier discussed has an infinite multipole series. The RSD enter in that the length of $k_3$ matters, which it does not in the no-RSD 3PCF.


We now evaluate the inverse FTs of the coefficients in equation (\ref{eqn:coadded_nok3_multis}) using equations (\ref{eqn:simple_inv_FT}) and (\ref{eqn:2d_transform}). Because the $k_1$ and $k_2$ dependence is separable, the inverse FTs can be further simplified to 1-D integral transforms of the power spectrum (closely related to Hankel transforms) defined by
\begin{align}
\xi^{[n]}_i &= \int \frac{k^2 dk}{2\pi^2}P(k)j_n(kr_i)\nonumber\\
\xi^{[n\pm]}_i &= \int \frac{k^2 dk}{2\pi^2}k^{\pm 1}P(k)j_n(kr_i);
\label{eqn:xi_fns}
\end{align}
subscript $i$ indicates the triangle side length $r_i$. Performing the inverse FT of the coefficients in equation (\ref{eqn:coadded_nok3_multis}) and adding our result equation (\ref{eqn:SQ2_k3dpdt_config}) for the $k_3$-dependent terms, we find the multipole coefficients of the redshift-space pre-cyclic 3PCF as
\begin{align}
&\ell = 0:\;b_1^3\bigg\{\frac{34}{21}\left[1+\frac{4}{3}\beta+\frac{1154}{1275}\beta^2+\frac{936}{2975}\beta^3+\frac{21}{425}\beta^4\right] \nonumber\\
&\;\;\;\;\;\;\;\;\;\;\;\; + \gamma \left[1+\frac{2}{3}\beta+\frac{1}{9}\beta^2\right]\bigg\}\xi^{[0]}_1 \xi^{[0]}_2 \nonumber\\
&\;\;\;\;\;\;\;\;\;\;\;\; + b_1^3(7\beta^2+3\beta^3)\kappa_0(r_1,r_2)\nonumber\\
&\ell = 1:\; -b_1^3\left[1+\frac{4}{3}\beta+\frac{82}{75}\beta^2+\frac{12}{25}\beta^3+\frac{3}{35}\beta^4\right]\nonumber\\%
&\;\;\;\;\;\;\;\;\;\;\;\;\;\times\bigg[\xi_1^{[1+]}\xi_2^{[1-]}+\xi_2^{[1+]}\xi_1^{[1-]}\bigg]\nonumber\\
&\;\;\;\;\;\;\;\;\;\;\;\; + b_1^3(7\beta^2+3\beta^3)\kappa_1(r_1,r_2)\nonumber\\
&\ell = 2:\;b_1^3\bigg\{\frac{8}{21}\left[1+\frac{4}{3}\beta+\frac{52}{21}\beta^2+\frac{81}{49}\beta^3+\frac{12}{35}\beta^4\right] \nonumber\\
&\;\;\;\;\;\;\;\;\;\;\;\; +\frac{32\gamma}{945}\beta^2\bigg\}\xi^{[2]}_1 \xi^{[2]}_2 + b_1^3(7\beta^2+3\beta^3)\kappa_2(r_1,r_2) \nonumber\\
&\ell = 3:\; -b_1^3\left[\frac{8}{75}\beta^2+\frac{16}{175}\beta^3+\frac{8}{315}\beta^4\right] \bigg[\xi_1^{[3+]}\xi_2^{[3-]}+\xi_2^{[3+]}\xi_1^{[3-]}\bigg] \nonumber\\
&\;\;\;\;\;\;\;\;\;\;\;\;\; + b_1^3(7\beta^2+3\beta^3)\kappa_3(r_1,r_2) \nonumber\\
&\ell = 4:\; b_1^3\left[-\frac{32}{3675}\beta^2+\frac{32}{8575}\beta^3+\frac{128}{11025}\beta^4\right]\xi^{[4]}_1 \xi^{[4]}_2 \nonumber\\
&\;\;\;\;\;\;\;\;\;\;\;\; + b_1^3(7\beta^2+3\beta^3) \kappa_4(r_1,r_2) \nonumber\\
&\ell \geq 5:\; b_1^3(7\beta^2+3\beta^3)\kappa_{\ell}(r_1,r_2) \nonumber\\
\label{eqn:config_space_model}
\end{align}
This equation is a primary result of this paper: the 3PCF equivalent of the
SCF99 model for the redshift-space line-of-sight-averaged bispectrum.

\subsection{Adding tidal tensor bias}
\label{subsec:tidal_tensor_bias}
Recent work (McDonald \& Roy 2009; Baldauf et al. 2012) suggests that galaxy bias models should incorporate a term proportional to the local tidal tensor; we denote this bias $b_{\rm t}$. The kernel generating the tidal tensor is simply $S_2(\hk_1\cdot\hk_2) = (2/3)P_2(x)$, and from Gil-Mar\'in et al. (2015) equations (C11) and (C12) it is evident that mapping $\tF(k_1,k_2;x)\to \tF(k_1,k_2;x)+(2\gamma'/3)P_2(x)$ in equation (\ref{eqn:model}), with $\gamma'\equiv 2 b_{\rm t}/b_1$, will incorporate tidal tensor bias. This replacement leads to a term $(2\gamma'/3)P_2(x)\mD_{\rm SQ1}(\beta,x)$ in equation (\ref{eqn:model}) and adds terms to equation (\ref{eqn:config_space_model}) as
\begin{align}
&\ell=0:\;\frac{16\beta^2\gamma'}{675}\xi^{[0]}_1 \xi^{[0]}_2\nonumber\\
&\ell=2:\;\frac{5}{2}\left(\frac{8}{15}+\frac{16\beta}{45}+\frac{344\beta^2}{4725}\right)\gamma' \xi^{[2]}_1 \xi^{[2]}_2\nonumber\\
&\ell=4:\;\frac{32\beta^2\gamma'}{525}\xi^{[4]}_1 \xi^{[4]}_2.
\label{eqn:tidal_multipoles}
\end{align}

For local Lagrangian biasing, one has (Baldauf et al. 2012; Chan, Socccimarro \& Sheth 2012)
\begin{align}
b_t = -\frac{2}{7}\left(b_1-1\right)
\label{eqn:b1_to_bt}
\end{align}
which for $b_1=2$ translates to $\gamma^{\prime} \approx -0.3$.

\subsection{Discussion of pre-cyclic results}
The combinations of $\xi^{[n]}$ and $\xi^{[n\pm]}$ that enter above are shown in Figure \ref{fig:xi_precyc}.  The $\kappa_{\ell}$ are shown in Figure \ref{fig:kappa_ell_precyc}, and the $\beta$-dependent coefficients above are shown in Figure \ref{fig:rsd_coeffs}, split out by bias coefficient. 

Figure \ref{fig:xi_precyc} shows that all of the combinations of $\xi^{[n]}$ and $\xi^{[n\pm]}$ entering the 3PCF have BAO features where either $r_1$ or $r_2$ nears the BAO scale of $100\Mpch$. We have not yet cyclically summed so the third triangle side $r_3$ has not yet entered the 3PCF and consequently there is no BAO structure related to it.  We will later find that after cyclic summing, it can add additional BAO structure to the 3PCF. In Figure \ref{fig:xi_precyc}, the monopole has a slight increment at the BAO scale and becomes negative as $r_1$ or $r_2$ exceeds the BAO scale (provided the other side remains below the BAO scale).  When both sides exceed the BAO scale the monopole becomes positive again.  This BAO feature is the analog of the BAO bump in the 2PCF; it comes from the BAO information in the galaxy density field. In particular, it is generated by the product of the velocity divergence and the density (SE15b; Bernardeau et al. 2002). Recalling the continuity equation, one can see that the velocity divergence will be in phase with the density, and so the $\ell=0$ BAO feature is in phase with the BAO in the density field. The $\ell=2$ term comes from gradients of the velocity divergence parallel to the velocity.  The $\ell=4$ term comes from coupling the 3PCF's intrinsic quadrupole $\ell=2$ with the RSD, which are a quadrupolar distortion.  

  
In contrast, in the dipole $(\ell = 1)$, there is a slight decrement for $r_1$ or $r_2$ just below the BAO scale and a slight increment as either side exceeds the BAO scale, with a zero crossing at the BAO scale.  This structure stems from gradients of the density field entering the 3PCF, as discussed in more detail in SE15b \S7.1 and S16a \S5.1. Given that the density field has a bump at the BAO scale, with positive slope prior to the bump, zero slope at the peak of the bump, and negative slope after the bump, one expects the gradient to cross zero at the bump and change sign from one side of the bump to the other.  The $\ell=3$ term comes from the 3PCF's intrinsic dipole $(\ell=1)$ coupled to the quadrupolar $(\ell=2)$ RSD, and thus has similar structure to $\ell=1$. Since the BAO structure in these odd pre-cyclic multipoles is generated by gradients of the density field, it is ``out-of-phase'' with the BAO structure in the even pre-cyclic multipoles.

The one-dimensional functions $\xi^{[0]}(r),\;\xi^{[1\pm]}(r)$, and $\xi^{[2]}(r)$ are shown in SE15a Figure 7, upper panel, and $\xi^{[1+]}(r)\xi^{[1-]}(r)$, the diagonal of the $\ell=1$ panel in the present work's Figure \ref{fig:xi_precyc}, is shown in SE15a Figure 7, lower panel. These plots further illustrate the difference in behavior between even and odd multipoles with respect to the BAO scale, in partciular that the $\ell=1$ BAO feature is rather different from that in $\ell=0$ and $\ell=2$, in being ``out-of-phase'' with the BAO in the density field because it stems from density gradients. These panels also again show that $\ell =1$ has both a larger overall amplitude and a somewhat stronger BAO feature than $\ell=0$ and $2$.  The lower panel of SE15a Figure 7 has been multiplied by a negative sign relative to the upper panel there and to the $\ell=1$ panel in Figure \ref{fig:xi_precyc} of the present work, because that panel only shows the term as it is enters the 3PCF (i.e. with a prefactor $-b_1^3$; see equation (\ref{eqn:config_space_model})).

As comparing Figure \ref{fig:kappa_ell_precyc} to Figure \ref{fig:xi_precyc} shows, the $\kappa_{\ell}$ are $\sim 0.1\%$ of the combinations of $\xi^{[n]}$ and $\xi^{[n\pm]}$ in equation (\ref{eqn:config_space_model}).  Furthermore, the $\kappa_{\ell}$ enter only beginning at $\oO(\beta^2)$ whereas the combinations of $\xi^{[n]}$ and $\xi^{[n\pm]}$ enter at order unity. Thus, while the $\kappa_{\ell}$ were the most involved to obtain, they are negligible in the 3PCF.  This conclusion is not unexpected given that only a small minority of terms in the bispectrum had $k_3$-dependence and produced $\kappa_{\ell}$ in configuration space. On the other hand, as already noted our decomposition into $k_3$-dependent and $k_3$-independent terms is not unique, so this comment is meant only to guide the intuition.

Figure \ref{fig:rsd_coeffs} shows the functions of $\beta$ entering equation (\ref{eqn:config_space_model}) split out by bias coefficient and including tidal tensor bias as given by equation (\ref{eqn:tidal_multipoles}). We stress how small the $\ell=3$ and $\ell=4$ terms are as compared to the $\ell=0,1$ and $2$ terms. This suppression was already evident by inspection of equations (\ref{eqn:config_space_model}) and (\ref{eqn:tidal_multipoles}) where $\ell=3$ and $\ell=4$ enter only at $\oO(\beta^2)$.  The suppression means that in practice these higher multipoles essentially do not affect the post-cyclic 3PCF.  

Second, the tidal tensor bias $\gamma^{\prime}$ only has a notable coefficient at $\ell=2$; tidal tensor biasing essentially does not affect the other pre-cyclic multipoles. We note also that the non-linear bias $b_2$ only enters $\ell = 0$ and $\ell=2$. We thus wish to consider comparisons between the $\ell=0$ and $\ell=2$, with linear and non-linear biasing, to $\ell=1$.  The $\ell=1$ coefficient in Figure \ref{fig:rsd_coeffs} (solid red line) is comparable to the relevant $\ell=0$ coefficients (black solid line and black dashed line), yet Figure \ref{fig:xi_precyc} shows that $\ell=0$ is intrinscially a factor of ten weaker than $\ell=1$. Consequently when the $\beta$-dependent coefficients are included, we expect $\ell=1$ to dominate over $\ell=0$ in the pre-cyclic 3PCF. This is shown in Figure \ref{fig:xi_precyc_w_prefacs}.

While intrinsically $\ell=1$ and $\ell=2$ have similar magnitudes, for $\ell=2$ only tidal tensor bias and linear bias are important, with non-linear biasing contributing very little. For $|\gamma^{\prime}| \sim \gamma$, the total $\ell=2$ coefficient can become rather small if $\gamma^{\prime}<0$. Meanwhile the $\ell=1$ coefficient is set purely by the solid red (linear bias) line in Figure \ref{fig:rsd_coeffs}; thus it can become significantly larger than the $\ell=2$ total coefficient. Conequently when the $\beta$-dependent pre-factors are included, $\ell=2$ can become very subdominant to $\ell=1$, as is indeed shown in Figure \ref{fig:xi_precyc_w_prefacs}.

Finally, while  Figure \ref{fig:xi_precyc} shows that $\ell=3$ is intrinsically fairly strong relative to $\ell=0$ through $2$, Figure \ref{fig:rsd_coeffs} shows its $\beta$-dependent pre-factor is much smaller, and so it makes a smaller contribution to the final pre-cyclic multipoles with these pre-factors included, as displayed in Figure \ref{fig:xi_precyc_w_prefacs}.  Meanwhile, while $\ell=4$ is of similar magnitude to $\ell=1$ and $2$, its $\beta$-dependent coefficient is much smaller and so it contributes little to the pre-cyclic 3PCF, as Figure \ref{fig:xi_precyc_w_prefacs} shows.

\subsection{Redshift space vs. real space}
\label{subsec:red_vs_real}
Confirming the finding of S16a, equation (\ref{eqn:config_space_model}) shows that at order $\beta$ RSD simply rescale the no-RSD 3PCF multipoles by a constant at each multipole. With no non-linear biasing $(\gamma=0)$, this $\oO(\beta)$-rescaling is multipole-independent and equal to $1+(4/3)\beta = 1.49$ for the S16a-adopted value of $\beta=0.37$ for the SDSS CMASS sample of Luminous Red Galaxies (LRGs).  For $\gamma\neq 0$ the $\ell=0$ recieves an additional contribution at $\oO(\beta)$. This contribution is not quite a simple rescaling, as the $\gamma$ term gets rescaled by $1+(2/3)\beta=1.25$ for $\beta=0.37$.  At $\oO(\beta)$ RSD thus slightly reduce the importance of non-linear biasing relative to linear biasing in the monopole.  

At $\oO(\beta^2)$ and again with $\gamma$ and $\gamma^{\prime}=0$, the rescaling begins to depend weakly on the multipole of the triangle's opening angle, with values $1.62$ for $\ell=0$, $1.64$ for $\ell=1$, and $1.83$ for $\ell=2$.  In the monopole, the importance of non-linear biasing relative to linear biasing is still reduced as compared to real-space. Interestingly, RSD also introduce information on the non-linear bias into $\ell=2$, in contrast to real space where non-linear bias entered only the monopole (the multipoles are with respect to the triangle's opening angle). RSD are a quadrupolar distortion so by angular momentum coupling it is expected that they introduce terms entering the real-space monopole into the redshift-space quadrupole.

With $\gamma$ and $\gamma^{\prime}=0$, the terms of $\oO(\beta^3)$ and higher constitute only $2.7\%$ of the total amplification of the monopole relative to real space, $3.9\%$ for the dipole. They are more important for the quadrupole, constituting $9.8\%$ of the total amplification relative to real space.  

Ignoring the $\kappa_{\ell}$ since they are extremely small as discussed previously, one can view RSD as a triangle-side-length independent renormalization of each multipole moment (but varying from multipole to multipole). For $\beta=0.37$ and $\gamma=0.3$, the rescaling of the monopole is $1.58$, the dipole $1.67$, and the quadrupole $1.93$. Including tidal tensor bias with $\gamma^{\prime}=-0.3$ does not alter the dipole and changes the rescalings for monopole and quadrupole by less than $1\%$.


\begin{figure*}
\centering
\includegraphics[width=.8\textwidth]{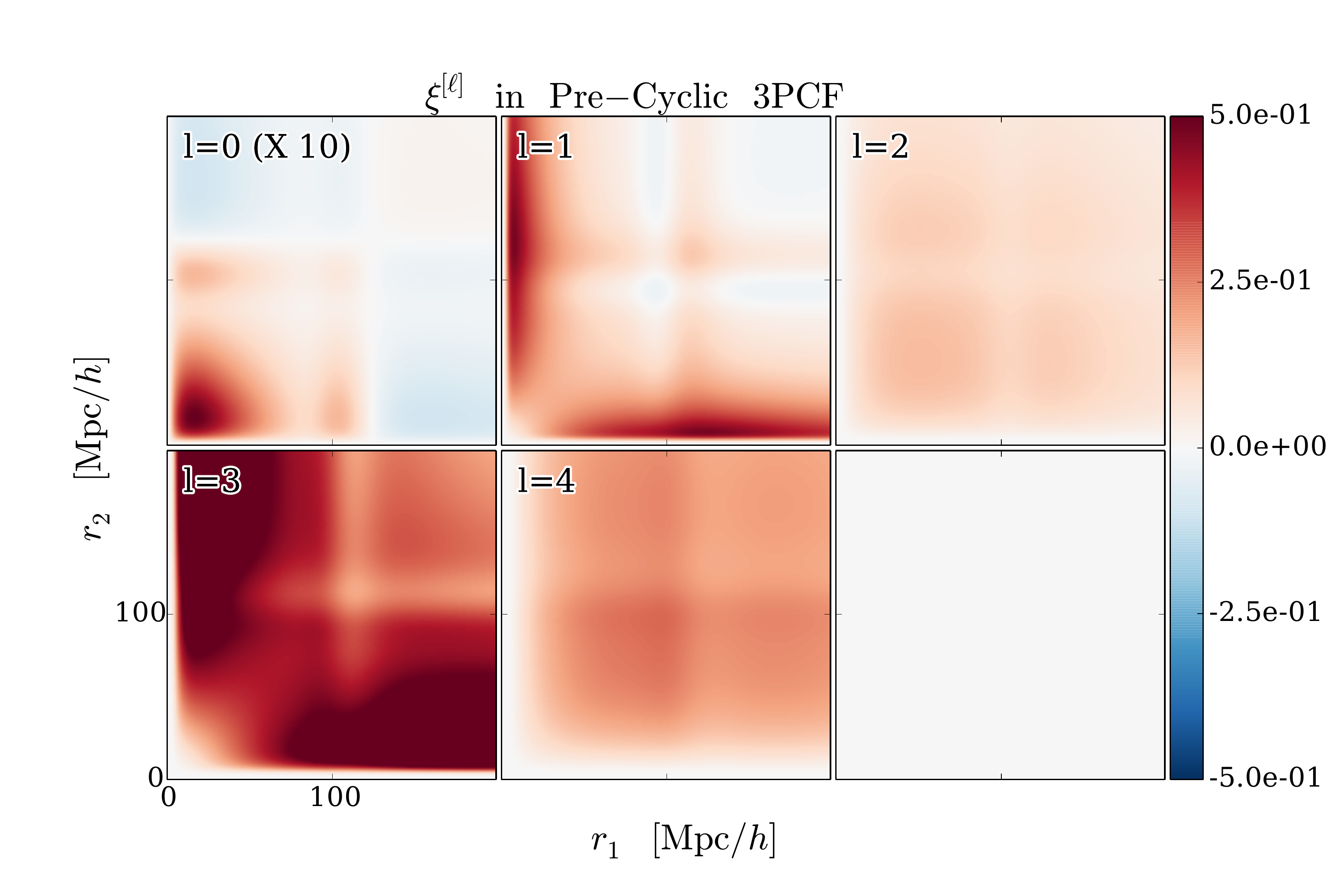}
\caption{The functions entering the pre-cyclic 3PCF as given in equation (\ref{eqn:config_space_model}), e.g. for $\ell=0$, $\xi^{[0]}(r_1)\xi^{[0]}(r_2)$, for $\ell=1$, $\xi^{[1+]}(r_1)\xi^{[1-]}(r_2)+{\rm symm.}$, for $\ell=2$, $\xi^{[2]}(r_1)\xi^{[2]}(r_2)$, for $\ell=3$, $\xi^{[3+]}(r_1)\xi^{[3-]}(r_2)+{\rm symm.}$, for $\ell=4$, $\xi^{[4]}(r_1)\xi^{[4]}(r_2)$, and for $\ell=5$, $\xi^{[5+]}(r_1)\xi^{[5-]}(r_2)+{\rm symm.}$  The $(\times\; 10)$ in the $\ell=0$ panel indicates that panel only has been multiplied by a factor of ten to be easily visible on the same colorbar as the other panels. We have weighted all panels by $r_1^2r_2^2/[10\Mpch]^4$ to take out the fall-off of the 3PCF on large scales, where it scales as $1/(r_1^2r_2^2)$. Recalling that $\zeta \sim \xi^2$ and that $\xi$ at $100\Mpch$ is of order $1\%$, the weighted 3PCF at this scale in $r_1$ and $r_2$ should be of order unity. It is somewhat less here because we are splitting its amplitude into multipoles. The rightmost lower panel has been intentionally left blank; there are only four $\xi^{[\ell]}$ entering the pre-cyclic 3PCF.}
\label{fig:xi_precyc}
\end{figure*}

\begin{figure*}
\centering
\includegraphics[width=.8\textwidth]{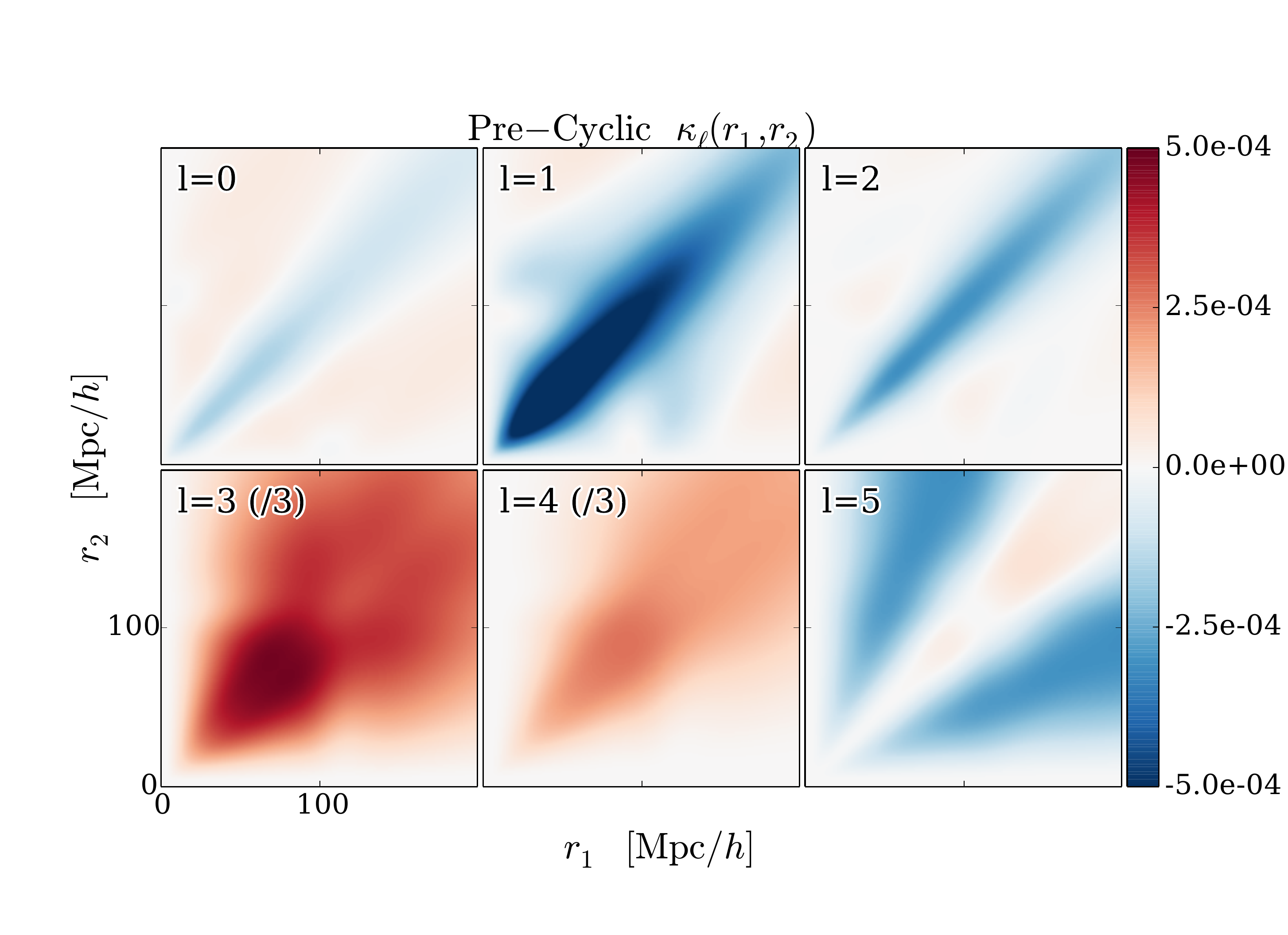}
\caption{The contribution of the $\kappa_{\ell}$ terms to the multipoles as in equation (\ref{eqn:config_space_model}). We have used the same weighting as in Figure \ref{fig:xi_precyc}. The key point of this plot is actually the colorbar: it is three orders of magnitude smaller than that for the $\xi^{[\ell]}$ combinations shown in Figure \ref{fig:xi_precyc}, so the $\kappa_{\ell}$ are a negligible contribution to the pre-cyclic 3PCF. The $\kappa_{\ell}$ do have interesting spatial structure, in particular a localized feature around the BAO in $\ell=1$ and a drop beginning at the BAO scale in $\ell=3$.  Nonethless, since the pre-cyclic 3PCF is dominated by the $\xi^{[\ell]}$, there is little chance of exploiting any additional BAO information introduced via the $\kappa_{\ell}$.}
\label{fig:kappa_ell_precyc}
\end{figure*}

\begin{figure}
\centering
\includegraphics[width=.55\textwidth]{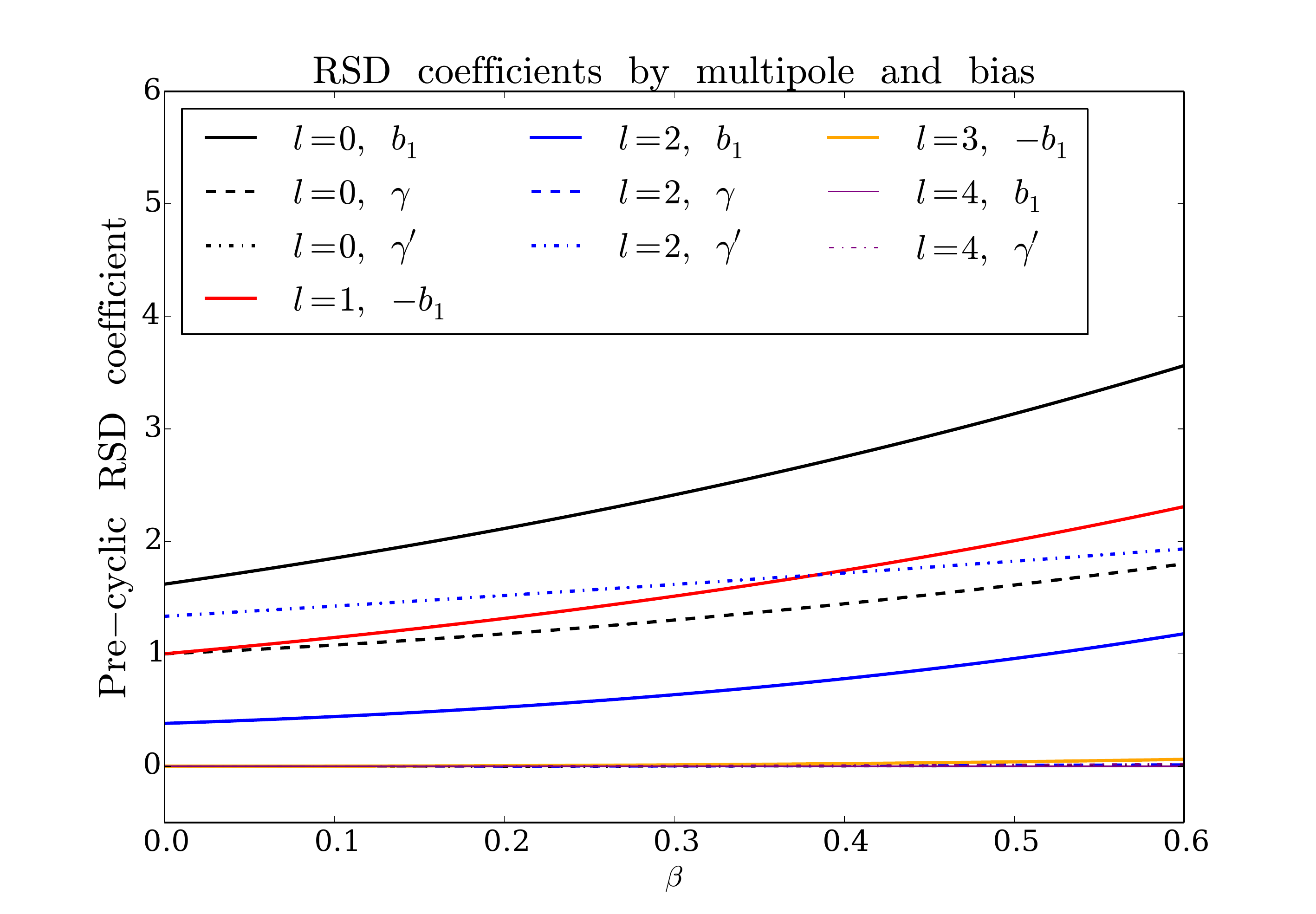}
\caption{The functions of $\beta$ entering the pre-cyclic 3PCF as in equation (\ref{eqn:config_space_model}).  We have split each mutlipole's coefficients by the bias on which they depend, since these biases are ultimately empirically determined.  We have multiplied the odd multipoles' coefficients by $-1$ to make the plot more compact. First, $\ell=3$ and $\ell=4$ are significantly sub-dominant to $\ell=0,\;1$ and $2$, as they enter only beginning at $\oO(\beta^2)$.  Second, the tidal tensor bias ($\gamma^{\prime}=b_t/b_1$) only enters significantly at $\ell=2$ and minimally affects the other multipoles. Finally, five curves of small magnitude overlap: the $\ell =0,\;\gamma^{\prime}$, $\ell=2,\;\gamma$, $\ell=3,\;b_1$, $\ell=4,\;b_1$, and $\ell = 4,\;\gamma^{\prime}$ are all clustered near zero amplitude throughout the plot.
}
\label{fig:rsd_coeffs}
\end{figure}

\begin{figure*}
\centering
\includegraphics[width=.9\textwidth]{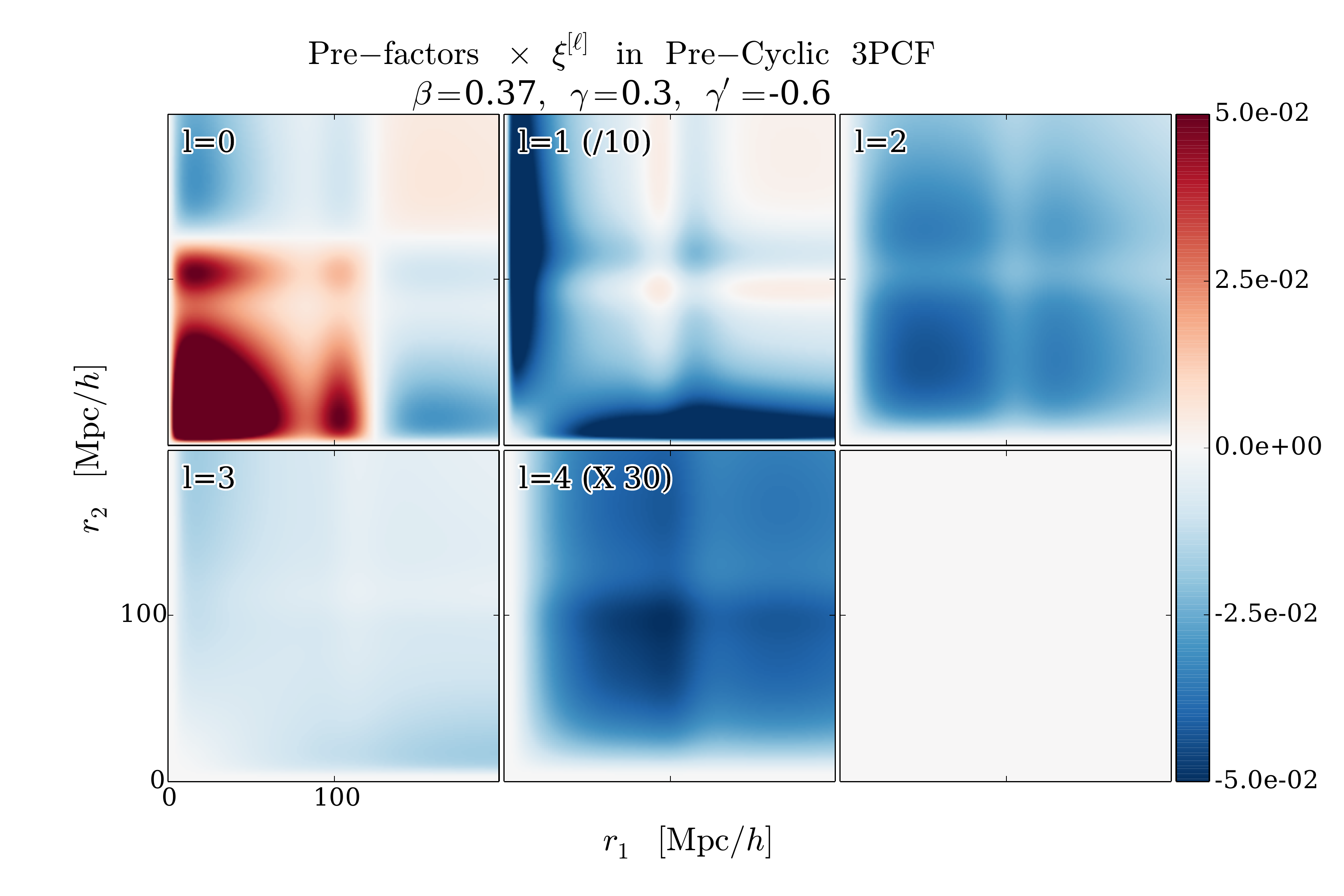}
\caption{Here we show Figure \ref{fig:xi_precyc} multiplied by the functions shown in Figure \ref{fig:rsd_coeffs} with the values of $\beta,\;\gamma$, and $\gamma^{\prime}$ indicated in the plot title; the motivation for these values is discussed in the main text.  As we have already noted, the $\kappa_{\ell}$ are negligible in the pre-cyclic 3PCF, so this Figure shows what is essentially the full pre-cyclic 3PCF. The weighting is the same as in Figure \ref{fig:xi_precyc}. We have divided the $\ell=1$ panel by ten and multiplied the $\ell=4$ panel by thirty so they are easily visible on the same colorbar as the other multipoles. This Figure shows that the $\ell=1$ term is by far the dominant one in the pre-cyclic 3PCF both in overall amplitude and BAO features. The $\ell=0$ and $\ell=2$ also contribute BAO information to the pre-cyclic 3PCF.
}
\label{fig:xi_precyc_w_prefacs}
\end{figure*}

\subsection{Further modeling non-linear structure formation and RSD}
\label{subsec:further_model}
In Gil-Mar\'in et al. (2015), the kernels $\tF$ and $\tG$ are promoted to ``effective'' kernels to incorporate non-linear structure formation and RSD (see their equations (D1) and (D2)). These effective kernels have the same angular dependence as $\tF$ and $\tG$ but have scale-dependence that individually rescales each multipole moment of $\tF$ and $\tG$.  This scale dependence is specified via functions (their equation (D6)) of nine free parameters for $\tF$ and nine free parameters for $\tG$ fixed from N-body simulations.  

Since the modifications of $\tF$ and $\tG$ affect the scale dependence but not the angular dependence, if desired they could be incorporated by using effective power spectra given by the product of the linear power spectrum and the functions of Gil-Mar\'in et al. (2015) equation (D6). Different effective power spectra would be required for each multipole moment of both $\tF$ and $\tG$, complicating the book-keeping, but the overall approach presented in the present work is sufficiently general to incorporate these kernels. The key property of the effective kernels that permits their use within our framework is that the radial and angular dependences remain separable. The fact that they alter only the radial dependence is additionally favorable for use within the present work's approach.

To model fingers-of-God from small-scale thermal velocities, Gil-Mar\'in et al. (2015) additionally add an overall pre-factor depending on the cosine $\mu_i$ of the angle between each of the three wavevectors $\vk_i$ and the line of sight (their equation (C15)). This term scales as $[1+\sum_i k_i \mu_i^2]^{-2}$ and so is not straightforwardly separable into radial and angular pieces. This term likely becomes relevant only on scales within a cluster, of order $10-20\Mpch$ or less.  We do not attempt to incorporate it in this work.

\section{Cyclic summing} 
\label{sec:cyc_sum}
The expressions presented thus far have been prior to the cyclic summing that accounts for the indistinguishability of triangle vertices in the 3PCF. The PT expressions above come from taking a particular galaxy to contribute each term in the product of three galaxy overdensities $\delta_{\rm g}$ that produces the 3PCF. Omitting tidal tensor biasing for simplicity, one has the galaxy bias model 
\begin{align}
\delta_{\rm g}(\vx_i) = b_1\delta_{\rm m}(\vx_i) + b_2\delta_{\rm m}^2(\vx_i),
\end{align}
where $\delta_{\rm m}(\vx_i) = \delta (\vx_i)+ \delta^{(2)}(\vx_i)$ is the matter overdensity, $\delta(\vx)_i$ the linear density field, and $\delta^{(2)}(\vx_i)$ is the second-order density field at a location $\vx_i$.  When forming the 3PCF, one takes the expectation value of $\delta_{\rm g}(\vx_i)\delta_{\rm g}(\vx_j)\delta_{\rm g}(\vx_k)$. In this expectation value, there will be particular combinations of densities that reach a given order in perturbation theory; for instance, at fourth (leading) order, one must have two linear density field and one second-order density field. Since the 3PCF is translation-invariant, the pre-cyclic calculation can be simplifed by assuming that the second-order density field is always contributed by the galaxy at $\vx_i$ and taking $\vx_i$ to be the origin of coordinates. 

However in reality we do not know which galaxy contributes which density field; each must have the chance to contribute all terms. Put another way, the full 3PCF must be symmetric under cyclic permutation of triangle side labels.  Consequently the predictions for the 3PCF presented in \S\ref{sec:RSD_model} must be cyclically summed around the triangle. This point is discussed in more detail in SE15a. Here we develop a computationally efficient approach to evaluating the required cyclic sums. One has the full 3PCF $\zeta$ as
\begin{align}
&\zeta(r_1,r_2;\hr_1\cdot\hr_2) = \sum_L\bigg[\zeta_{{\rm pc},L}(r_1,r_2)P_L(\hr_1\cdot\hr_2) \nonumber\\
&+ \zeta_{{\rm pc},L}(r_2,r_3)P_L(\hr_2\cdot\hr_3) + \zeta_{{\rm pc},L}(r_3,r_1)P_L(\hr_3\cdot\hr_1)\bigg],
\label{eqn:cyc_sum_defn}
\end{align}
where subscript $pc$ denotes ``pre-cyclic'' and refers to the model of \S\ref{sec:RSD_model}. The multipole moments of the post-cyclic 3PCF are then 
\begin{align}
\zeta_l(r_1,r_2)=\frac{2l+1}{2}\int \frac{d\Omega_1d\Omega_2}{16\pi^2}P_l(\hr_1\cdot\hr_2)\zeta(r_1,r_2;\hr_1\cdot\hr_2),
\label{eqn:cyc_sum_reprojection}
\end{align}
where the factor of $(2l+1)/2$ enters because the Legendre basis is orthogonal but not orthonormal.

A direct way to evaluate this integral, which we used in SE15a, is to define $r_1, r_2$, and the opening angle cosine $\mu_{12}=\hr_1\cdot\hr_2$.  For each such triplet, $r_3$ can be computed by the law of cosines. This procedure scales as $N_r^2N_{\mu}$, where $N_r$ is the number of grid points in $r_1$ or $r_2$ used and $N_{\mu}$ the number of grid points in $\mu_{12}$.

Given that the angular dependence is ultimately integrated out, we consider whether the angular part of the cyclic summing can be done analytically. In particular, we will show that such an approach allows a cyclic summing that scales as $N_{\rm bins}^2$, with $N_{\rm bins}$ the number of radial bins used for $r_1$ and $r_2$, with no dependence on $N_{\mu}$ at all.

We first discuss the cyclic sums resulting from the terms in equation (\ref{eqn:config_space_model}) generated by $k_3$-independent components of the bispectrum.  These cyclic sums are all of the form $\xi^{[L]}(r_1)\xi^{[L]}(r_2)P_L(\hr_1\cdot\hr_2)+{\rm cyc.}$ or $\xi^{[L+]}(r_1)\xi^{[L-]}(r_2)P_L(\hr_1\cdot\hr_2) +{\rm cyc.}$ The cyclic piece replaces either $r_1$ or $r_2$ with $r_3=|\vr_1-\vr_2|$ and $\hr_1$ or $\hr_2$ with $\hr_3$. The functions involving $\vr_3$ can be rewritten as integral transforms of the power spectrum using the definition (\ref{eqn:xi_fns}); without loss of generality we concentrate on the product $\xi^{[L]}(r_1)\xi^{[L]}(r_3)$:
\begin{align}
&\xi^{[L]}(r_1)\xi^{[L]}(r_3)P_L(\hr_1\cdot\hr_3) \nonumber\\
&=\xi^{[L]}(r_1) \int\frac{k^2dk}{2\pi^2}j_L(k|\vr_1 - \vr_2|) P(k) P_L(\hr_i\cdot\reallywidehat{\vr_1- \vr_2}).
\label{eqn:half_real_half_fourier}
\end{align}
Focusing on the $\vr_1$ and $\vr_2$-dependent pieces above and using the spherical harmonic addition theorem to expand the Legendre polynomial, we find
\begin{align}
&j_L(k|\vr_1 - \vr_2|) P_L(\hr_1\cdot\reallywidehat{\vr_1 - \vr_2})\nonumber\\
&=\frac{4\pi}{2L+1}j_L(k|\vr_1 - \vr_2|) \sum_{M=-L}^L Y_{LM}(\hr_1)Y_{LM}^*(\reallywidehat{\vr_1- \vr_2})
\label{eqn:interm_step}
\end{align}
Using the identity (\ref{eqn:sep_id}) proven in Appendix B to rewrite the righthand side of equation (\ref{eqn:interm_step}) as a separated product of spherical Bessel functions and spherical harmonics in $\vr_1$ and $\vr_2$, inserting what results in equation (\ref{eqn:half_real_half_fourier}), and simplifying, we obtain
\begin{align}
&\xi^{[L]}(r_1)\xi^{[L]}(r_3)P_L(\hr_1\cdot\hr_3)= \xi^{[L]}(r_1)\int\frac{k^2 dk}{2\pi^2} P(k) \frac{(4\pi)^2}{2L+1}\nonumber\\
&\times \sum_{M=-L}^L Y_{LM}^*(\hr_1)\sum_{L_1 M_1} \sum_{L_2 M_2} i^{L_2-L_1+L} j_{L_1}(kr_1)j_{L_2}(kr_2)\nonumber\\
&\times \mathcal{C}_{L_1 L_2 L}
\left(\begin{array}{ccc}
L_{1} & L_2 & L\\
0 & 0 & 0
\end{array}\right)
\left(\begin{array}{ccc}
L_{1} & L_2 & L\\
M_1 & M_2 & M
\end{array}\right)\nonumber\\
&\times Y_{L_1 M_1}^*(\hr_1)Y_{L_2 M_2}^*(\hr_2),
\label{eqn:interm_step2}
\end{align}
where the $2\times3$ matrices are Wigner 3j-symbols describing the addition of total angular momenta (top row) and spin angular momenta (bottom row). We have defined $\mathcal{C}_{l_1 l_2 l_3} = \sqrt{[(2l_1+1)(2l_2+1)(2l_3+1)]/(4\pi)}$.

Wishing now to obtain the coefficients of the above in the basis of Legendre polynomials in $\hr_1\cdot\hr_2$, we integrate equation (\ref{eqn:interm_step2}) against $(2l+1)/(16\pi^2)P_l(\hr_1\cdot\hr_2)d\Omega_1 d\Omega_2$, using the spherical harmonic addition theorem to expand the Legendre polynomial and then invoking 3j-symbols and orthogonality for the integrals over respectively $d\Omega_1$ and $d\Omega_2$. We then sum the spin-dependent 3j-symbols that result over all spins using NIST Digital Library of Mathematical Functions (DLMF) 34.3.18, finding 
\begin{align}
&\xi^{[L]}(r_1)\xi^{[L]}(r_3)P_L(\hr_1\cdot\hr_3)= \sum_l \zeta_l(r_1, r_2)P_l(\hr_1\cdot\hr_2)\nonumber\\
&{\rm with}\nonumber\\
&\zeta_l(r_1, r_2) =\frac{4\pi}{2L+1} \xi^{[L]}(r_1)\sum_{L_1}i^{l+L-L_1}\mathcal{C}^2_{L l L_1}\left(\begin{array}{ccc}
L & l & L_1\\
0 & 0 & 0
\end{array}\right)^2\nonumber\\
&\times\int\frac{k^2 dk}{2\pi^2} P(k)j_{L_1}(kr_1)j_l(kr_2).
\end{align}

We now incorporate radial binning, which integrates $r_1$ and $r_2$ to be within bins $S_1$ and $S_2$.  For the binning in $r_2$, we can simply replace $j_l(kr_2)$ above with its bin-averaged value $\bar{j}_l(S_2; k)$ as defined in SE15b equation (70).  For the binning in $r_1$, we require
\begin{align}
\Phi_{LL_1}(S_1; k)\equiv \frac{4\pi}{V(S_1)}\int^{S_1+\Delta}_{S_1-\Delta} r_1^2 dr_1 \xi^{[L]}(r_1)j_{L_1}(kr_1),
\label{eqn:psi_defn}
\end{align}  
where $\Delta$ is half the bin width and $V(S_1)$ is the volume of the bin centered at $S_1$. Our final binned result, denoted by a bar, is thus
\begin{align}
&\bar{\zeta}_l(r_1, r_2) =\frac{4\pi}{2L+1} \sum_{L_1}i^{l+L-L_1}\mathcal{C}^2_{L l L_1}\left(\begin{array}{ccc}
L & l & L_1\\
0 & 0 & 0
\end{array}\right)^2\nonumber\\
&\times\int\frac{k^2 dk}{2\pi^2} P(k)\Phi_{LL_1}(S_1; k)\bar{j}_l(S_2;k).
\end{align}
Thus once the $\Phi_{LL_1}$ and $\bar{j}_l$ are computed for all desired bins and multipoles, we need only do a 1-D integral transform of $N_{\rm bins}^2$ combinations of these functions.

We now discuss the application of this cyclic summing strategy to the terms in $\kappa_{\ell}(r_1,r_2)$. From equations (\ref{eqn:kappa_ell}-\ref{eqn:ftens_defn}) we see that pre-cyclically the $r_1$ and $r_2$-dependence of $\kappa_{\ell}(r_1,r_2)$is simply $j_{\ell}(kr_1)j_{\ell}(kr_2)$, with $P_{\ell}(\hr_1\cdot\hr_2)$ the angular dependence implied by a given multipole $\ell$. This dependence has exactly the same form as that of the pre-cyclic terms already discussed, and so the approach outlined in that context may be employed for cyclically summing the $\kappa_{\ell}$ as well. As earlier noted, even pre-cyclically the $\kappa_{\ell}$ involve all $\ell$, and arbitrarily high pre-cyclic $L$ can project onto a given post-cyclic $\ell$, even if $\ell$ is small.  Consequently in principle we should cpmpute an infinite number of pre-cyclic $\kappa_{L}$ and then project onto each post-cyclic $\ell$.  However in practice we find the post-cyclic result is well converged after using pre-cyclic $L$ up to 6, and for safety in the results presented here we use pre-cyclic $L$ up to $12$.  

Finally, for the plots presented in this work, we did not implement the cyclic summing scheme presented above but used the slower, direct method discussed at the beginning of this section.  While this direct method is not optimally efficient, it was adequate for predicting the 3PCF from a single set of cosmological parameters in a reasonable amount of time. Future work will be implementing the more efficient approach presented here.

\clearpage
\begin{figure*}
\centering
\includegraphics[width=.9\textwidth]{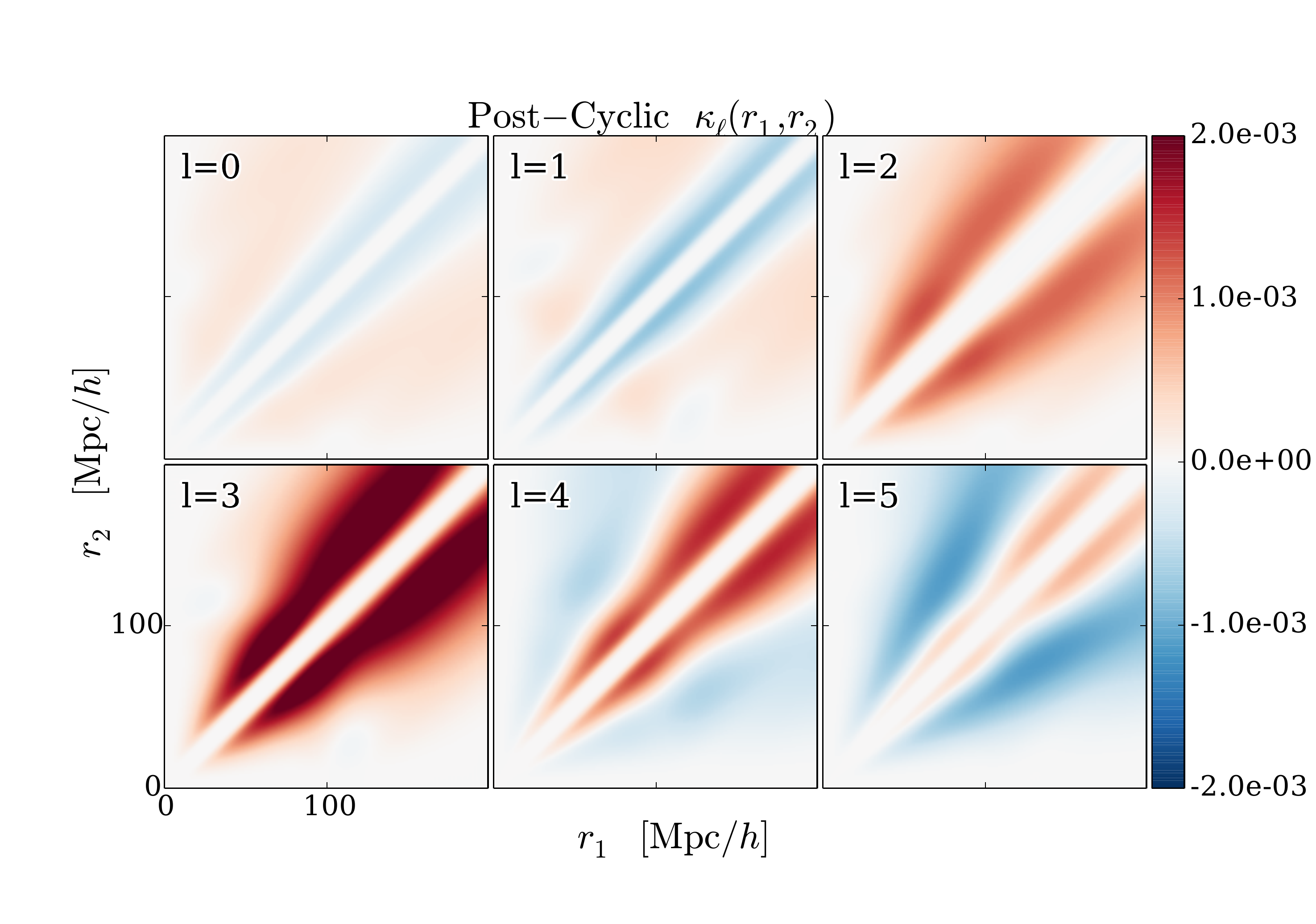}
\caption{The multipoles $\kappa_{\ell}$ of the $k_3$-dependent terms after cyclic summing and reprojection onto the Legendre basis. We have weighted by $r_1^2r_2^2/[10\Mpch]^4\exp\left[-[12\;\Mpch]^2/(r_1-r_2)^2\right]$ to take out the fall-off of the 3PCF on large scales, where it behaves as  $1/(r_1^2r_2^2)$. This weighting also suppresses the diagonal. The post-cyclic $\kappa_{\ell}$ again have some BAO features (most prominently in $\ell=1$). However, their amplitude is at least two orders of magntiude smaller than that of the post-cyclic 3PCF due to the $\xi^{[\ell]}$, shown in Figures \ref{fig:pre_to_post}-\ref{fig:post_cyc_3pcf}.  Consequently the $\kappa_{\ell}$ are negligible in the post-cyclic 3PCF, as we expected given that this was so in the pre-cyclic 3PCF. The $\kappa_{\ell}$ contribute post-cyclically at all multipoles; we have chosen to show only the first six, but the contribution of the higher-$\ell$ $\kappa_{\ell}$ to higher multipoles of the post-cyclic 3PCF is similarly negligible. See \S\ref{subsec:kappa_ell_pre_to_post} for further discussion.
}
\label{fig:kappa_ell_post_cyc}
\end{figure*}
\clearpage

\clearpage
\begin{figure*}
\centering
\includegraphics[width=.8\textwidth]{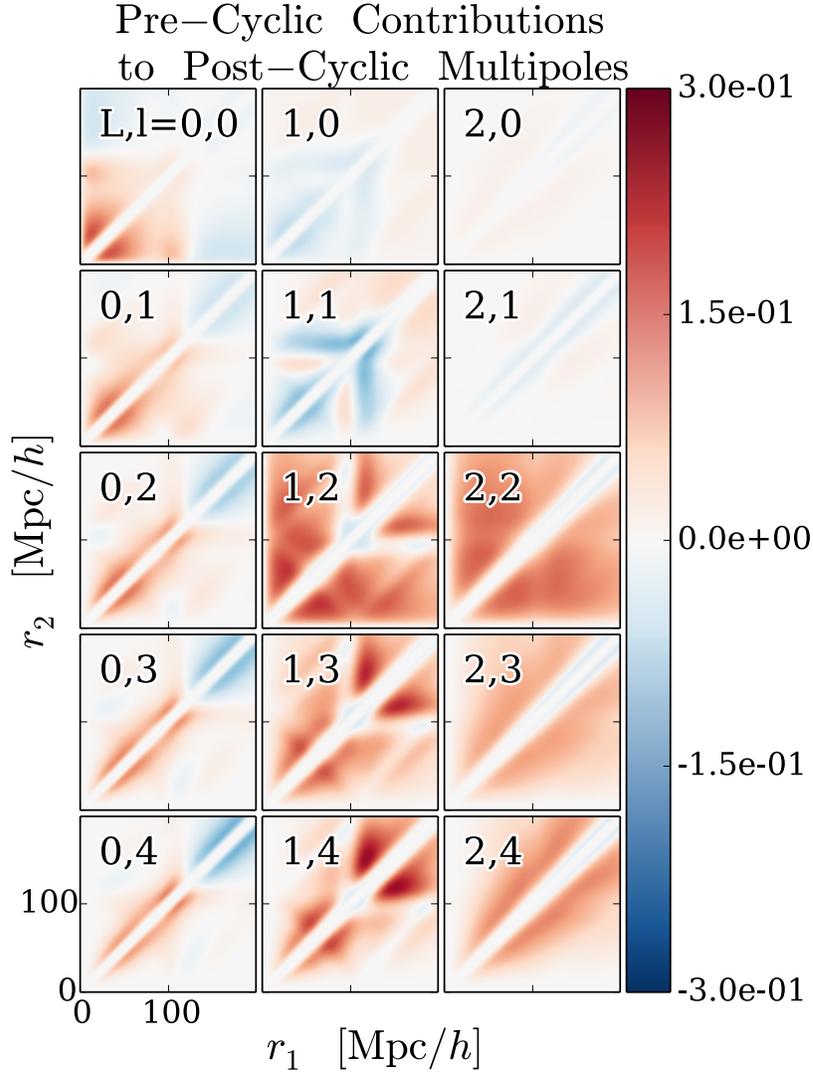}
\caption{The contribution of each pre-cyclic multipole $L$ to each post-cyclic multipole $l$. In each panel, the ordered pair is pre-cyclic, post-cyclic.  We only display pre-cyclic multipoles $0$ through $2$ as the pre-cyclic $L=3$ and $4$ have much smaller magnitudes.  The pre-cyclic multipoles $L\geq 5$ are even smaller than those for $L=3$ and $4$, as they are sourced purely by the $\kappa_{\ell}$ (see equation (\ref{eqn:config_space_model})). To obtain the full post-cyclic 3PCF at each multipole, one would add these panels across rows (and also include the negligible higher-$L$ pre-cyclic contributions we do not show).  The key point of this Figure is that the $L=1$ and $2$ pre-cyclic multipoles dominate the post-cyclic 3PCF.  The $L=0$ pre-cyclic multipole also contributes to the post-cyclic multipoles but is generally weaker around the BAO scale, save for in the $l=0$ post-cyclic mutlipole where it is the dominant BAO-scale contribution. The $L=1$ pre-cyclic multipole contributes to the $l=2$ post-cyclic multipole as much as the $L=2$ pre-cyclic multipole does; furthermore, the BAO structure in the post-cyclic $l=2$ panel comes primarily from the pre-cyclic $L=1$, as is also the case in the higher-$l$ post-cyclic multipoles. Overall, the pre-cyclic $L=1$ multipole is the dominant source of BAO information in the post-cyclic 3PCF. These points are further discussed in \S\ref{subsec:kappa_ell_pre_to_post}.
}
\label{fig:pre_to_post}
\end{figure*}

\clearpage
\begin{figure*}
\centering
\includegraphics[width=.8\textwidth]{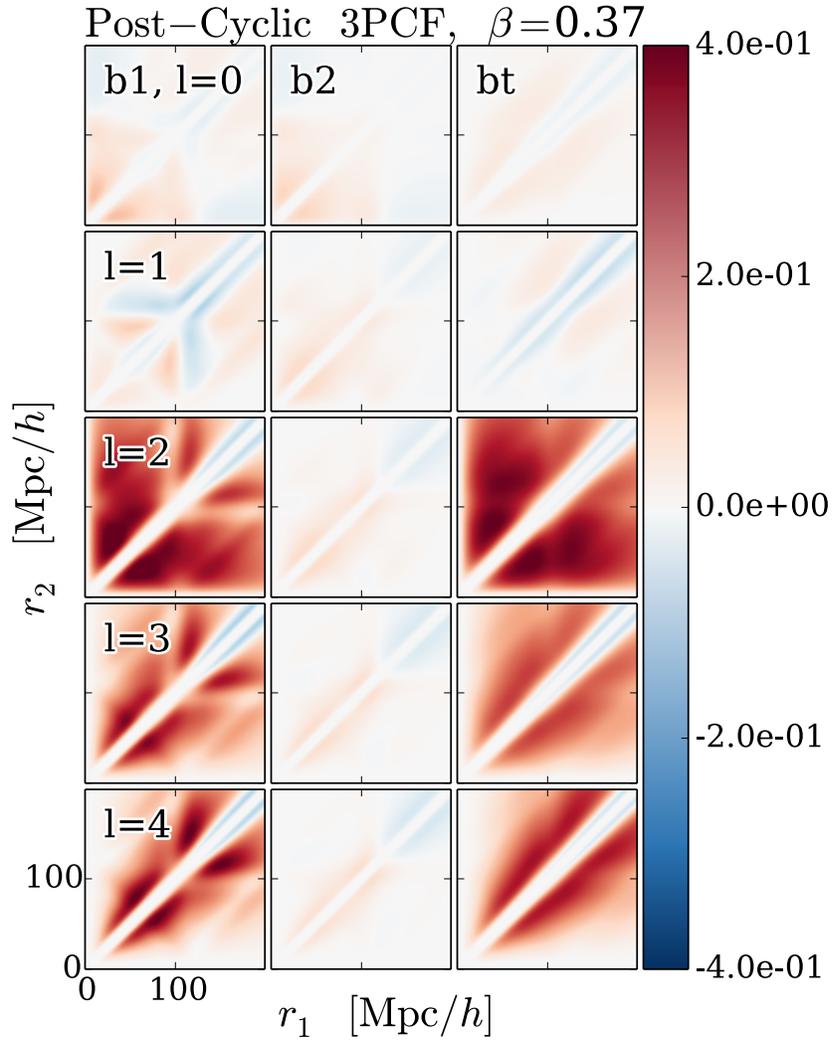}
\caption{The multipoles of the post-cyclic 3PCF split out by bias coefficient and excluding the contribution of $\kappa_{\ell}$. To obtain the full post-cyclic 3PCF (exlcuding the negligible $\kappa_{\ell}$) one would add across rows weighted by values of $b_1,\;b_2$, and $b_t$.  Note the weakness of the $b_2$-dependent multipoles relative to the $b_1$ and $b_t$-dependent multipoles. Note also the prominent BAO feature in the $b_1,\;l=1$ panel. Also evident is the $l=2$ multipole's dominance in amplitude; this is because it receives a significant contribution both from the pre-cyclic $L=1$ and $L=2$, as shown in Figure \ref{fig:pre_to_post}. Finally, observe the similiarity of the $l=3$ and $l=4$ panels within each bias coefficient; this similarity continues within each bias coefficient for the higher multipoles we do not show. It occurs because pre-cyclically there is structure only in $L=0,1$ and $2$ to a good approximation. See \S\ref{subsec:3pcf_split_by_bias} for further discussion.
}
\label{fig:no_kappa_post_cyc}
\end{figure*}

\clearpage
\begin{figure*}
\centering
\includegraphics[width=.9\textwidth]{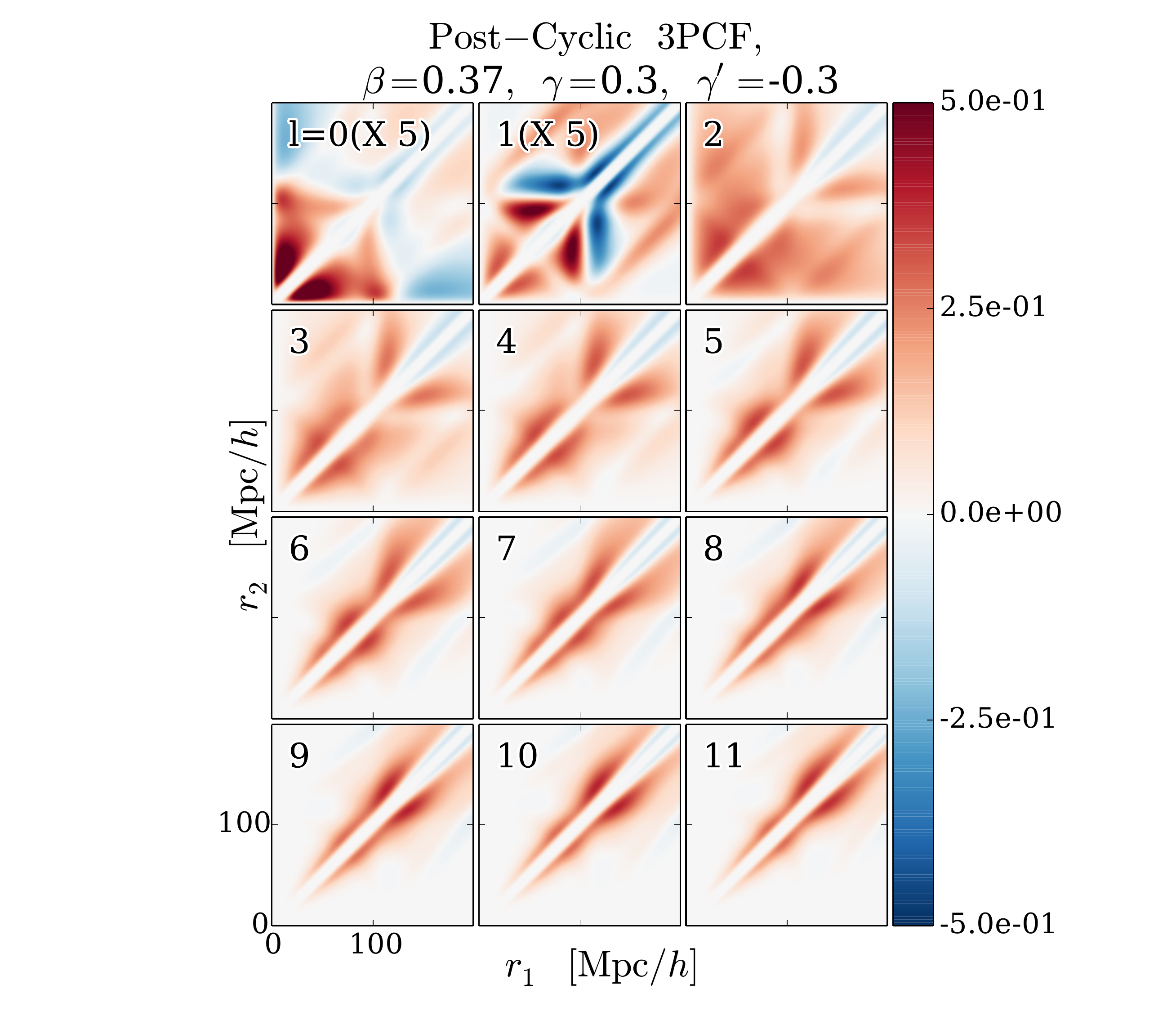}
\caption{The full post-cyclic 3PCF, having chosen values of the biases $b_2$ and $b_t$ ($b_1$ scales out) and included the terms in $\kappa_{\ell}$.  The $l=0$ and $l=1$ panels have been multiplied by five to show them clearly on the same colorbar as the other panels.  We note the strong BAO features in the $l=1$ panel, and to a lesser extent also in the $l=0$ panel.  Further, the multipoles $l \geq 3$ are all  similar to each other; this traces back to the pre-cyclic 3PCF's having structure only in $L=0,1$, and $2$ to a good approximation.  This Figure shows that measuring the first four multipoles of the 3PCF likely extracts most of the BAO information, and that the higher multipoles are not highly independent from each other. As $l$ rises the multipoles are more and more dominated by squeezed triangles near the diagonals of these panels. These points are further discussed in \S\ref{subsec:total_3pcf_fixed_bias}.
}
\label{fig:post_cyc_3pcf}
\end{figure*}

\clearpage
\begin{figure*}
\centering
\includegraphics[width=1.\textwidth]{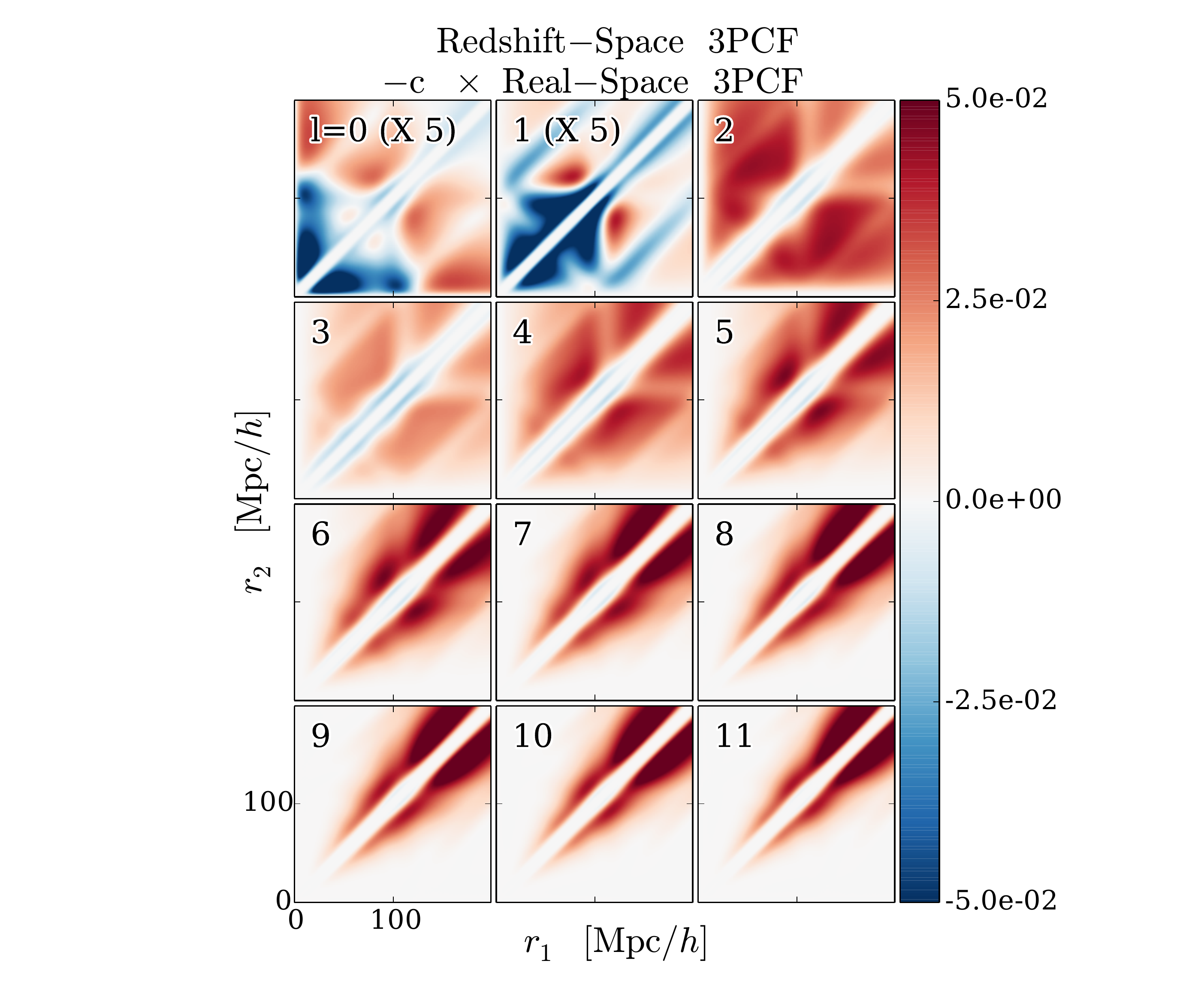}
\caption{The redshift-space 3PCF minus a constant times the real-space 3PCF (we have chosen $c = 1.8$).  The colorbar of this Figure is reduced by a factor of ten relative to that of Figure \ref{fig:post_cyc_3pcf}. The small magnitude of these panels in comparison to the redshift-space 3PCF shows that the redshift-space 3PCF is roughly a constant rescaling of the real-space 3PCF. The constant will depend on the values of $\beta,\;b_1,\;b_2$, and $b_t$. We have chosen to subtract a constant multiple of the real-space 3PCF rather than divide by it to avoid division by zero when the real-space 3PCF has zero-crossings. This rescaling is further discussed in \S\ref{subsec:redshift_space_rescale}.}
\label{fig:redshift_minus_real}
\end{figure*}

\clearpage
\begin{figure*}
\includegraphics[width=1.\textwidth]{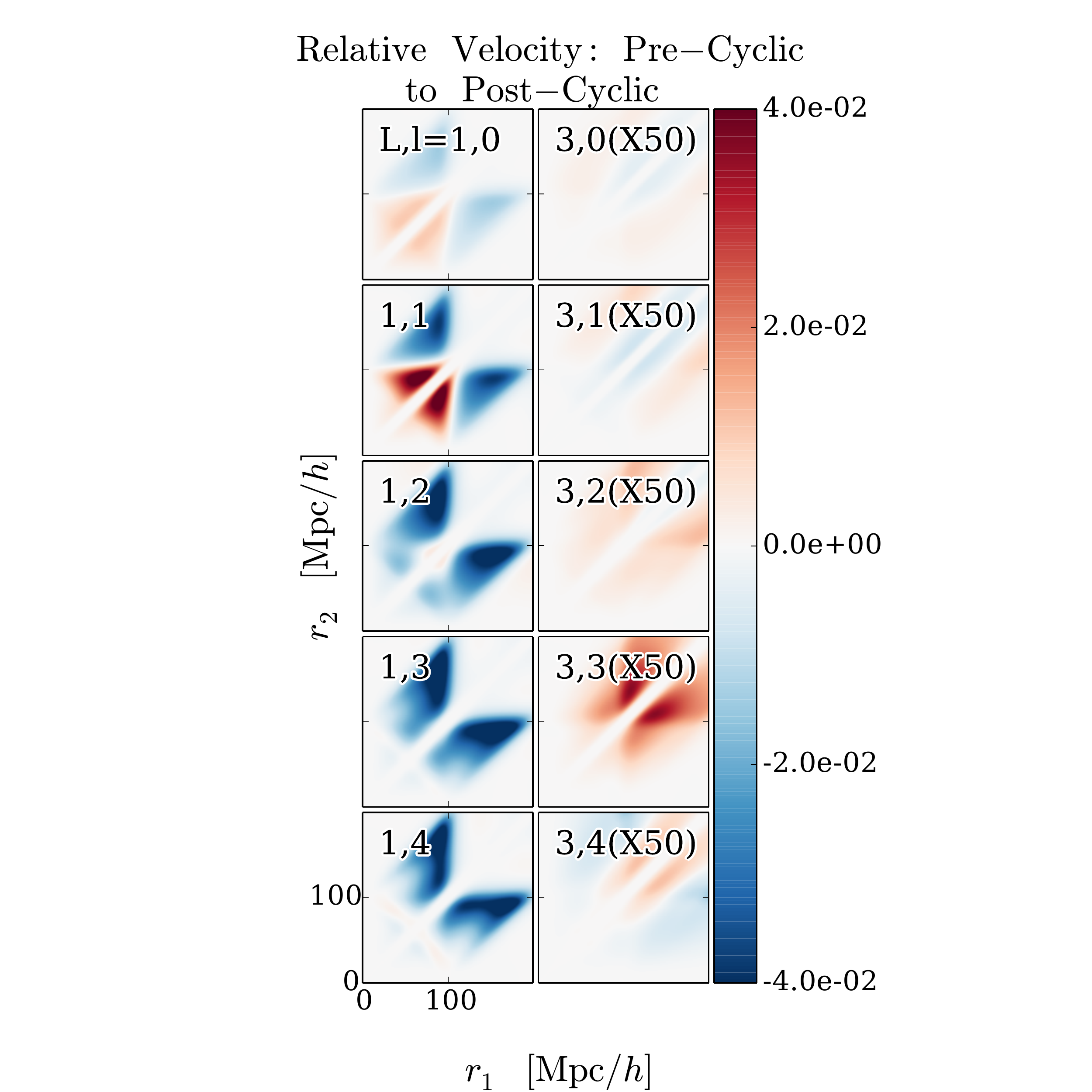}
\caption{The post-cyclic relative velocity contributions to the redshift-space 3PCF, split by the pre-cyclic multipole (either $L=1$ or $3$) generating the contribution.  The contribution of the pre-cyclic $L=3$ multipole is negligible; each panel in that column has been multiplied by fifty.  The $L=1$ column duplicates the result of SE15a (Figure 9, third column) for the relative velocity contribution to the real-space 3PCF, up to a constant pre-factor depending on $\beta$. We note the prominent BAO features in the $1,1$ panel here, further discussed in \S\ref{sec:rv_contrib} and in SE15a.}
\label{fig:rv_pre_to_post}
\end{figure*}

\section{Discussion}
\label{sec:disc}
We now discuss the results of cyclically summing the 3PCF model presented in \S\ref{sec:RSD_model}. In all post-cyclic plots, the diagonal has been suppressed by a factor $\exp\left[-[12\Mpc]^2/(r_1-r_2)^2\right]$ since it becomes large yet we expect is poorly modeled by perturbation theory. In particular, on the diagonal $r_1 = r_2$, meaning for a squeezed triangle $r_3 =0$.  The hierarchical ansatz $\zeta\sim \xi(r_1)\xi(r_2)+{\rm cyc.}$ (Peebles \& Groth 1977), with the linear correlation function $\xi\sim 1/r^2$, suggests the 3PCF has terms that look like $1/r_3^2$ which become large as $r_3$ becomes small. However, for any two galaxies close to each other (say $\lesssim 20\Mpch$), perturbation theory likely provides an inadequate description of the density field. We thus prefer to avoid showing the predictions in this regime, nor were they used in our previous analysis S16a.

\subsection{Post-cyclic results: $k_3$-dependent contributions and multipole coupling}
\label{subsec:kappa_ell_pre_to_post}
Figure \ref{fig:kappa_ell_post_cyc} shows the post-cyclic $\kappa_{\ell}(r_1, r_2)$.  Figure \ref{fig:pre_to_post} shows the contribution of each pre-cyclic multipole, not including the $\kappa_{\ell}$, to the post-cyclic multipoles zero through four.  Comparing the colorbars of these two Figures shows that the $\kappa_{\ell}$ are still negligible post-cyclically ; the $\kappa_{\ell}$ are two orders of magnitude less than the non-$\kappa_{\ell}$ contributions. Further,  Figure \ref{fig:pre_to_post} includes the $\beta$-dependent coefficients of equation (\ref{eqn:config_space_model}), whereas Figure \ref{fig:kappa_ell_post_cyc} does not.  Since $\kappa_{\ell}$ enters only beginning at $\oO(\beta^2)$, it will be even further suppressed relative to the $\xi^{[n]}$ and $\xi^{[n\pm]}$ contributions shown in Figure \ref{fig:pre_to_post}. 

Consequently the behavior of $\kappa_{\ell}$ shown in Figure \ref{fig:kappa_ell_post_cyc} will not significantly impact the 3PCF.  Nonetheless, the BAO do manifest in the $\kappa_{\ell}$. A very slight BAO feature in the $\ell=0$ and $\ell=1$ panels, most visible where one side is near the BAO scale of $100\Mpch$  and the other side becomes near zero.  There is also a slight bump in the $\ell=2,3$ and $4$ panels when both sides become near the BAO scale.  In the $\ell=5$ panel, there is a slight narrowing of the red band along the diagonal relative to the blue just where both sides reach the BAO scale.

We now turn to the coupling between pre-cyclic and post-cyclic multipoles excluding the $\kappa_{\ell}$, as shown in Figure \ref{fig:pre_to_post}. This figure shows the contribution of a pre-cyclic multipole $L$ to a post-cyclic multipole $l$. Here and in what follows, we always use $L$ to denote a pre-cyclic multipole and $l$ to denote a post-cyclic multipole. The main point this Figure conveys is that the $\ell=1$ and $\ell=2$ pre-cyclic multipoles are the most important contribution to the post-cyclic multipoles.  This can be seen from the strength of the $1,x$ and $2,x$ panels relative to the $0,x$ panels, where $x$ ranges from $0$ to $4$.  All of the strongest post-cyclic panels trace back to the pre-cyclic $\ell=1$ and $2$ multipoles. 

In more detail, this plot should be read across the row to determine the full post-cyclic multipole behavior, so comparing relative magnitudes within a row shows what pre-cyclic multipole dominates the post-cyclic multipole shown in that row.  For all but the top (post-cyclic $\ell=0$) row, the $L=1$ and $L=2$ contribution to a given post-cyclic $l$ dominates the $L=0$ contribution.  For $L=l=0$, the pre-cyclic $L=0$ contribution dominates the pre-cyclic $L=1$ and $2$, as one might expect because this is the diagoal coupling.  

Regarding diagonal coupling, it is slightly surprising that $1,2$ coupling is as strong as the $2,2$ coupling: pre-cyclic $L=1$ enters the post-cyclic $l=2$ just as much as pre-cyclic $L=2$ does.  More interestingly, given the smoothness of the $2,2$ panel, it is the $1,2$ panel that actually sets the spatial structure of the total post-cyclic $l=2$ multipole.  This point is important because the $L=1$ pre-cyclic multipole has rather strong BAO features, and its significant coupling to the post-cyclic $l=2$ is the reason for these features in the measured $l=2$ moment of the 3PCF.  

There are strong BAO features in the $1,1$ panel: the pre-cyclic $L=1$ sources a significant peak and then trough in the post-cyclic $l=1$ when one side or the other crosses the BAO scale.  This feature is the reason for the strong peak and trough in the compressed basis 3PCF measurement of the SDSS DR12 CMASS sample presented in S16a (in that work, see Figures 4 (mock catalogs) and 6 (data), $l=1$ panel).  Figure \ref{fig:pre_to_post} also shows that the BAO information in higher post-cyclic multipoles (e.g. $l=3$ and $l=4$) is due to the pre-cyclic $L=1$ multipole. 

Finally, the panels seem to converge as the post-cyclic $l$ becomes much greater than the pre-cyclic $L$ (these panels are the lower half-triangle of Figure \ref{fig:pre_to_post}). The contribution of a given pre-cyclic multipole is roughly the same to all higher post-cyclic multipoles.  This result is not unexpected: the functions of $\vr_3=|\vr_1+\vr_2|$ entering the cyclic sum generically produce coupling of a given pre-cyclic multipole to all post-cyclic multipoles, but there is nothing preferred about any post-cyclic $l$ not equal to the pre-cyclic $L$.  This comment applies equally when the pre-cyclic multipole is less than the post-cyclic multipole, as illustrated by the strong similarity of the $2,0$ and $2,1$ panels.

\subsection{3PCF split by bias coefficient}
\label{subsec:3pcf_split_by_bias}
Figure \ref{fig:no_kappa_post_cyc} shows the post-cyclic 3PCF split by the three bias coefficients $b_1$ (linear bias), $b_2$ (non-linear bias), and $b_t$ (tidal tensor bias) and by multipole. We have fixed $\beta = 0.37$ for these plots to match the $\beta$ expected for the SDSS DR12 CMASS sample used in S16a (see \S5.2 of that work).  

To obtain the total 3PCF at a given multipole from this Figure, one would sum across each row with the desired values of $b_1$, $b_2$, and $b_t$.  In $l=0$, $b_t$ does not have a visible BAO feature, save for perhaps a small shift from near zero to negative along the diagonal as both sides exceed the BAO scale.  Both $b_1$ and $b_2$ go from positive to negative as each side crosses the BAO scale.  Thus, any BAO feature in the measured 3PCF monopole comes primarily from linear and non-linear biasing.  

In $l=1$, the linear bias dominates the non-linear and tidal bias, and especially with regard to the BAO features. This dominance is unsurprising since pre-cyclically $b_2$ and $b_t$ have no $l=1$ behavior, entering only at even multipoles; in contrast, $b_1$ enters $l=1$ pre-cyclically. There is a strong BAO feature in the linear bias panel; a peak as either side approaches the BAO, a zero at the BAO scale, and a trough as either side crosses the BAO scale.  

In $l=2$, the dominant terms are the linear and tidal tensor biasing.  This dominance is expected because the non-linear bias only enters $l=2$ pre-cyclically at $\oO(\beta^2)$, whereas the linear bias and tidal tensor bias enter $l=2$ pre-cyclically at order unity.  This pre-cyclic behavior is in turn because the linear bias and the tidal bias both encode dynamics through respectively the $F_2$ and $S_2$ kernels, which have  $l=2$ terms (indeed, the $S_2$ kernel involves solely $l=2$).  In contrast, the non-linear biasing only encodes the square of the linear density, encodinng no dynamics and having no angular structure pre-cyclically. In general, this lack of dynamical information and intrinsic angular structure is the reason that all the multipoles of the non-linear bias look roughly the same.  Regarding BAO features, the $l=2$ linear bias has a decrement as either side becomes equal to the BAO scale with the other side greater than the BAO scale, and a slight bump when one side crosses the BAO scale while the other side is less than the BAO scale.  The tidal tensor bias has a slight decrement as either side crosses the BAO scale.

Within each bias, the $l=3$ and $l=4$ multipoles look similar.  This occurs for the same reasons that these panels look similar at a given pre-cyclic $L$ in Figure \ref{fig:pre_to_post}: there is essentially no intrinsic angular structure at $L>2$ in the pre-cyclic 3PCF (save for small contributions at $\oO(\beta^2)$), and so there is no preferred post-cyclic multipole $l$ when $l>2$. The tidal tensor bias panels are rather smooth for $l=3$ and $l=4$; this is because the $S_2$ kernel generating them has no scale dependence, in contrast to the $F_2$ kernel which involves $(k_1/k_2+k_2/k_1)$ and consequently produces stronger localized features in configuration space, such as those entering $l=3$ and $l=4$ in the linear bias panels.  As we have already noted, the non-linear bias panels look similar at all multipoles because this bias term has essentially no intrinsic angular dependence (it enters at order unity only in the pre-cyclic $L=0$).  

\subsection{Total 3PCF for fixed biases}
\label{subsec:total_3pcf_fixed_bias}
Figure \ref{fig:post_cyc_3pcf} shows the full multipoles of the post-cyclic 3PCF, with $\beta = 0.37, \gamma = 0.3$, and $\gamma^{\prime} = -0.3$.  $\gamma = 0.3$ is chosen to be consistent with both S16a and with the Gil-Mar\' in et al. (2015) measurement of the SDSS DR11 bispectrum.  $\gamma^{\prime}$ is chosen to follow the relation for local Lagrangian biasing (\ref{eqn:b1_to_bt}). These panels are generated by coadding the panels across each row in Figure \ref{fig:no_kappa_post_cyc} with appropriate coefficients.  

The panels look somewhat similar for $l\geq 2$, with the diagonal increasingly dominating as $l$ rises, due to the fact that higher multipoles place more weight on squeezed triangles where the opening angle is zero. In particular, around $\mu=1$ the $P_l$ have series $1-[l(l+1)/2](1-\mu)$, meaning a more severe drop in weight away from $\mu=1$ the larger $l$ is.  As already discussed, these $\mu\sim1$ triangles dominate the diagonal; hence the greater the weight placed on them as $l$ increases, the greater the diagonal relative to off-diagonal 3PCF.  Consequently we expect multipoles with $l\gg 2$ may not add much information to a 3PCF analysis. They will also likely have low signal in the off-diagonal radial bins actually used in a 3PCF analysis, as the compressed higher multipoles in S16a indeed show (compare the $l=2$ panel in S16a Figures 4 (mocks) and 6 (data) with the higher $l$ panels in Figures 5 (mocks) and 7 (data)).

Of the $l=0,1$ and $2$ panels, the $l=2$ dominates the $l=0$ and $1$. This dominance is because $l=2$ receives equal contributions from the pre-cyclic $L=1$ and $L=2$, whereas the post-cyclic $l=0$ and $l=1$ receive contributions mostly only from respectively pre-cyclic $L=0$ and $L=1$.  The creasing in $l=2$ on the downward-sloping line from $r_2 = 100\Mpch$ to $r_1=100\Mpch$ is a BAO feature occuring when the two sides $r_1+r_2=100\Mpch$.  Further, there are upward-sloping creases at larger scales, from $r_1,r_2=(100\Mpch,0)$ to $r_1,r_2=(200\Mpch,100\Mpch)$, and analagously with $r_1$ and $r_2$ switched.  These are BAO features where $r_2+r_2=200\Mpch$, twice the BAO scale.  Finally, there are BAO features when either $r_1$ or $r_2$ equals the BAO scale and the other side is larger than the BAO scale, visible as the horizontal and vertical white stripes in the panel. We can confirm these features are due to the BAO by making the same plots using the ``no-wiggle'' power spectrum, a power spectrum computed from the transfer function of Eisenstein \& Hu (1998) which models the growth of matter perturbations in the presence of baryons correctly but does not have any BAO features. 

Both $l=0$ and $l=1$ also have BAO features.  Comparing with Figure \ref{fig:pre_to_post}, one can see that these features trace back respectively to the $0,0$ and $1,1$ panels of that Figure, as expected. It is important to notice that in $l=1$ the BAO feature is an increment slightly below the BAO scale and a decrement slightly above it (bright red to bright blue as $r_1$ or $r_2$ crosses the BAO scale), with zeros at the BAO scale. This structure occurs because the $l=1$ BAO feature is sourced by gradients of the density field. In particular, the density Green's function has positive slope just below the BAO scale and negative slope just beyond it, with a zero-slope peak at the BAO scale (see Figure 6 of Slepian \& Eisenstein 2016).\footnote{The density Green's function is the late-time density generated by an initial Dirac-delta function perturbation in an otherwise homogeneous universe, and is the inverse Fourier transform of the transfer function.} In contrast, the BAO feature in $l=0$ comes from the density field itself, and hence has a peak as either triangle side crosses the BAO scale.  Both $l=0$ and $l=1$ also have diagonal creases with the same equations as those noted for $l=2$, and for the same reasons.

\subsection{Redshift-space rescaling of real-space 3PCF}
\label{subsec:redshift_space_rescale}
In S16a, we found that a real-space model with no treatment of RSD could fit the compressed 3PCF (one triangle side is integrated out over an annulus set by the other) well, with $\chi^2/{\rm d.o.f.}\approx 1$.  Motivated by this finding, we consider whether the redshift-space 3PCF is roughly a constant rescaling of the real-space 3PCF. Were this so, RSD would be highly degenerate with the value of the linear bias: in particular, a higher bias value could mimic the effect of RSD. Indeed, the bias values found in S16a were somewhat higher than those reported in Gil-Mar\'in et al. (2015) for a similar sample.

Figure \ref{fig:redshift_minus_real} shows that the redshift-space 3PCF is roughly a constant rescaling of the real-space 3PCF.  To avoid dividing by zero where the real-space 3PCF has zero crossings, we illustrate this point by subtracting the real-space 3PCF multiplied by a constant, $c$, from the redshift-space 3PCF.  The smallness of the difference as compared to the redshift-space 3PCF shown in Figure \ref{fig:post_cyc_3pcf} shows that deviations from a pure rescaling (in which case Figure \ref{fig:redshift_minus_real} would be unfiormly zero) are small.  We used $c=1.8$, chosen by eye to minimize the difference, and of the same order as the pre-cyclic rescalings listed in \S\ref{subsec:red_vs_real}. It is not unexpected that this roughly-optimal $c$ differs from the pre-cyclic rescalings, as pre-cyclically there is multipole-dependence in the rescalings and cyclic summing mixes the multipoles. In particular, the quadrupole is the strongest pre-cyclically, so $c$ should be closer to its pre-cyclic rescaling (1.93) than to that for the monopole (1.58) or dipole (1.67). Nonetheless, $c$ should be lower than the pre-cyclic quadrupole rescaling, as we indeed find, because these other pre-cyclic multipoles do pull the post-cyclic rescaling down.

\section{Relative velocity bias}
\label{sec:rv_contrib}
\subsection{Importance of the relative velocity bias}
Recent work by Tseliakhovich \& Hirata (2010) has shown that there is a relative velocity $\vbc$ between baryons and dark matter generated by their different behavior on sub-sound-horizon scales prior to decoupling, and that this relative velocity can add a term to the galaxy bias model (Yoo, Dalal \& Seljak 2011; hereafter YDS11).  Physically, the relative velocity field's coherence length is the BAO scale, as shown in Figure 2 of SE15a. Different regions of the Universe with different initial density perturbations will have different relative velocities.  The relative velocity's root-mean square is of order $10\%$ of the circular velcoity or velcoity dispersion of the smallest ($\sim 10^6\;M_{\odot}$) dark matter halos acquiring baryons at $z\sim50$, when the first galaxies are believed to form.  The relative velocity increases the baryons' kinetic energy in the dark matter halos' rest frames and making it more difficult for these halos to bind baryons.  

Galaxy formation may thus be modulated between different large-scale patches of the Universe, and as YDS11, SE15a, and Blazek, McEwen \& Hirata (2016; hereafter BMH16) show, this modulation can alter the galaxy 2PCF and 3PCF.  The galaxy bias model used in YDS11 and SE15a was based on SPT in real space and included a second-order term in $\vbc$. The galaxy bias model of BMH16, based on Lagrangian Perturbation Theory, includes an additional, advection term in $\vbc$. This advection term enters because galaxies' formation is affected by the relative velocity field at their Lagrangian position (i.e. where they formed), not at their present-day locations. Their subsequent advection is set by the matter field's velocity, which is highly correlated with the relative velocity, as SE15a Figure 2 shows (compare the dark matter velocity and the relative velocity). Consequently the advection term can significantly enhance the relative velocity's impact on the 2PCF.  YDS11 and SE15a found that the relative velocity bias can shift the BAO bump in the 2PCF used as a cosmic distance scale, and BMH16 shows the advection term greatly increases this shift for a given value of the relative velocity bias, $b_v$.  

\subsection{Deriving the redshift-space relative velocity 3PCF contribution}
The BMH16 advection term as it enters the bias model already involves three powers of the linear density field, and thus can only enter the 3PCF via terms with six powers of the linear density field. Thus, while the advection term enters the 2PCF in terms with four powers of the linear density field and for consistency should be added to the model for the power spectrum/2PCF used in YDS11 and SE15a, the bispectrum/3PCF model of these works need not be modified; it is still complete at fourth order in the linear density field. Nonetheless, these models are in real space; here we derive the relative velocity's contribution to the 3PCF in redshift space.

We begin with the product of the relative velocity's square and the density fields on a particular triangle with vertices $\vx,\;\vx+\vr_1$, and $\vx+\vr_2$:

\begin{align}
&\vbc^2(\vx,z)\delta(\vx+\vr_1,z)\delta(\vx+\vr_2,z)=-\int\frac{d^3\vk d^3\vk'}{(2\pi)^6} \frac{d^3\vk_1  d^3\vk_2}{(2\pi)^6} \nonumber\\
&\times e^{-i(\vk + \vk')\cdot \vx} e^{-i\vk_1\cdot(\vx +\vr_1)} e^{-i\vk_1\cdot(\vx +\vr_2)}(\hk\cdot \hk') \nonumber\\
& \times T_{\rm vbc}(k,z) T_{\rm vbc}(k',z) \tdelta_{\rm pri}(\vk) \tdelta_{\rm pri}(\vk') \tdelta(\vk_1,z)\tdelta(\vk_2,z);
\label{eqn:starting_point}
\end{align}
for the relative velocity's square we used SE15a equation (14). $T_{\rm vbc}$ is the relative velocity transfer function defined in SE15a equation (12), which converts a primordial density field $\tdelta_{\rm pri}$ into the Fourier-space relative velocity at redshift $z$; $\tdelta(\vk,z)$ is the linear density field at redshift $z$, such that $\tdelta(\vk, z) = T_{\rm m}(k,z)\tdelta_{\rm pri}(\vk)$, with $T_{\rm m}$ the matter transfer function.  

To obtain the leading-order contribution to the redshift-space 3PCF, we need only convert two of the four density fields in equation (\ref{eqn:starting_point}) to redshift space; at leading order converting only two agrees with converting all four. Similarly, to obtain the leading-order 3PCF contribution we need not convert the real-space positions to redshift space. We choose to convert the late-time density fields $\tdelta(\vk_1)$ and $\tdelta(\vk_2)$, employing the Kaiser (1987) formula 
\begin{align}
\tdelta_{\rm s}(\vk_i)=(1+\beta (\hk_i\cdot\hz)^2)\tdelta(\vk_i),
\label{eqn:kaiser_conversion}
\end{align}
where $\hz$ is the direction of the line of sight. Using this relation and averaging over all lines of sight by integrating against $d\Omega_z/(4\pi)$, we find
\begin{align}
&\left<\tdelta_{\rm s}(\vk_1)\tdelta_{\rm s}(\vk_2)\right>_z=\nonumber\\
& \bigg[1+\frac{2}{3}\beta +\frac{2}{15}\beta^2 +\frac{2}{15}\beta^2(\hk_1\cdot\hk_2)^2\bigg]\tdelta(\vk_1)\tdelta(\vk_2).
\label{eqn:los_avg}
\end{align}

We now insert equation (\ref{eqn:los_avg}) in equation (\ref{eqn:starting_point}) and take the expectation value over many realizations of the Gaussian Random Field density using Wick's Theorem, dropping the disconnected part.  This procedure sets $\vk=-\vk_1$ and $\vk'=-\vk_2$, causing the arguments of all the exponentials in $\vx$ to sum to zero; it also converts pairs of density fields to power spectra. We find

\begin{align}
&\left<\vbc^2(\vx)\delta(\vx+\vr_1)\delta(\vx+\vr_2)\right> = -2\int \frac{d^3\vk_1 d^3\vk_2}{(2\pi)^6} e^{-i\vk_1\cdot\vr_1}e^{-i\vk_2\cdot\vr_2} \nonumber\\
&\times T_{\rm vbc}(k_1)T_{\rm vbc}(k_2)P_{\times}(k_1)P_{\times}(k_2)\nonumber\\
&\times \bigg[ \bigg(1+\frac{2}{3}\beta+\frac{11\beta^2}{75} \bigg)P_1(\hk_1\cdot\hk_2) + \frac{4\beta^2}{75}P_3(\hk_1\cdot\hk_2)\bigg];
\label{eqn:last_fourier_space_step}
\end{align}
$P_{\times}(k)$ is a cross-power spectrum between a late-time linear density field and a primordial density field, defined via $(2\pi)^3\delta_{\rm D}^{[3]}(\vk+\vk')P_{\times}(k)=\left< \tdelta(\vk) \tdelta_{\rm pri}(\vk')\right>$

Using a theorem proven in SE15a Appendix A, the inverse Fourier transforms can be evaluated to give the relative velocity contribution to the redshift-space 3PCF, where we have now included the appropriate bias factors:
\begin{align}
&\zeta_{\rm vbc}(r_1,r_2;\hr_1\cdot\hr_2)=\nonumber\\
& 2b_1^2b_v \bigg[\bigg(1+\frac{2}{3}\beta+\frac{11\beta^2}{75} \bigg)  \xi^{[1]}_{\rm vbc}(r_1) \xi^{[1]}_{\rm vbc}(r_2) P_1(\hr_1\cdot\hr_2) \nonumber\\
&+ \frac{4\beta^2}{75} \xi^{[3]}_{\rm vbc}(r_1) \xi^{[3]}_{\rm vbc}(r_2) P_3(\hr_1\cdot\hr_2)\bigg]
\label{eqn:rv_contrib}
\end{align}
with 
\begin{align}
\xi^{[n]}_{\rm vbc}(r)\equiv \int\frac{k^2dk}{2\pi^2} j_n(kr)T_{\rm vbc}(k)P_{\times}(k);
\end{align}
the overall minus sign in equation (\ref{eqn:last_fourier_space_step}) was canceled by a factor of $(-1)^n$, with $n=1$ or $3$, entering the theorem proven in SE15a.

\subsection{Observational signatures}
Figure \ref{fig:rv_pre_to_post} shows the relative velocity contribution split by the two pre-cyclic multipoles that it produces in redshift space. This Figure shows that the dominant relative velocity contribution on scales below $100\Mpch$ is in the dipole, and is generated by the pre-cyclic dipole.  The higher multipoles also have relative velocity signatures from the pre-cyclic dipole at larger scales, but we expect the signal-to-noise on these larger scales will be low.  

While the pre-cylic relative velocity octopole ($\ell=3$) also contributes some distinctive structure post-cyclically, these post-cyclic amplitudes are greatly suppresssed relative to the pre-cyclic $\ell=1$ contribution. The right column of Figure \ref{fig:rv_pre_to_post} is multiplied by a factor of fifty to be visible on the same colorbar as the left column.

Figure \ref{fig:pre_cyclic_ell_one_fns} shows the functions entering the pre-cyclic 3PCF at $\ell=1$, including the relative velocity contributions.  The essential point of this Figure is that the spatial dependence of these functions is very different and consequently the relative velocity should not be degenerate with the linear bias.  While the ampliutde of the pre-cyclic dipole $(\ell=1)$ contribution appears small relative to that of the pre-cyclic octopole $(\ell=3)$ in this panel, there are pre-factors involving powers of $\beta$ that enter the 3PCF.  The octopole enters only at $\oO(\beta^3)$ while the dipole enters at $\oO(1)$, so the dipole is actually the important relative velocity contribution. It is the dashed red curve Figure \ref{fig:pre_cyclic_ell_one_fns}, and its distinctive causal structure (non-zero only within the sound horizon of $100\Mpch$) is discussed in detail in SE15a.

\begin{figure}
\centering
\includegraphics[width=.5\textwidth]{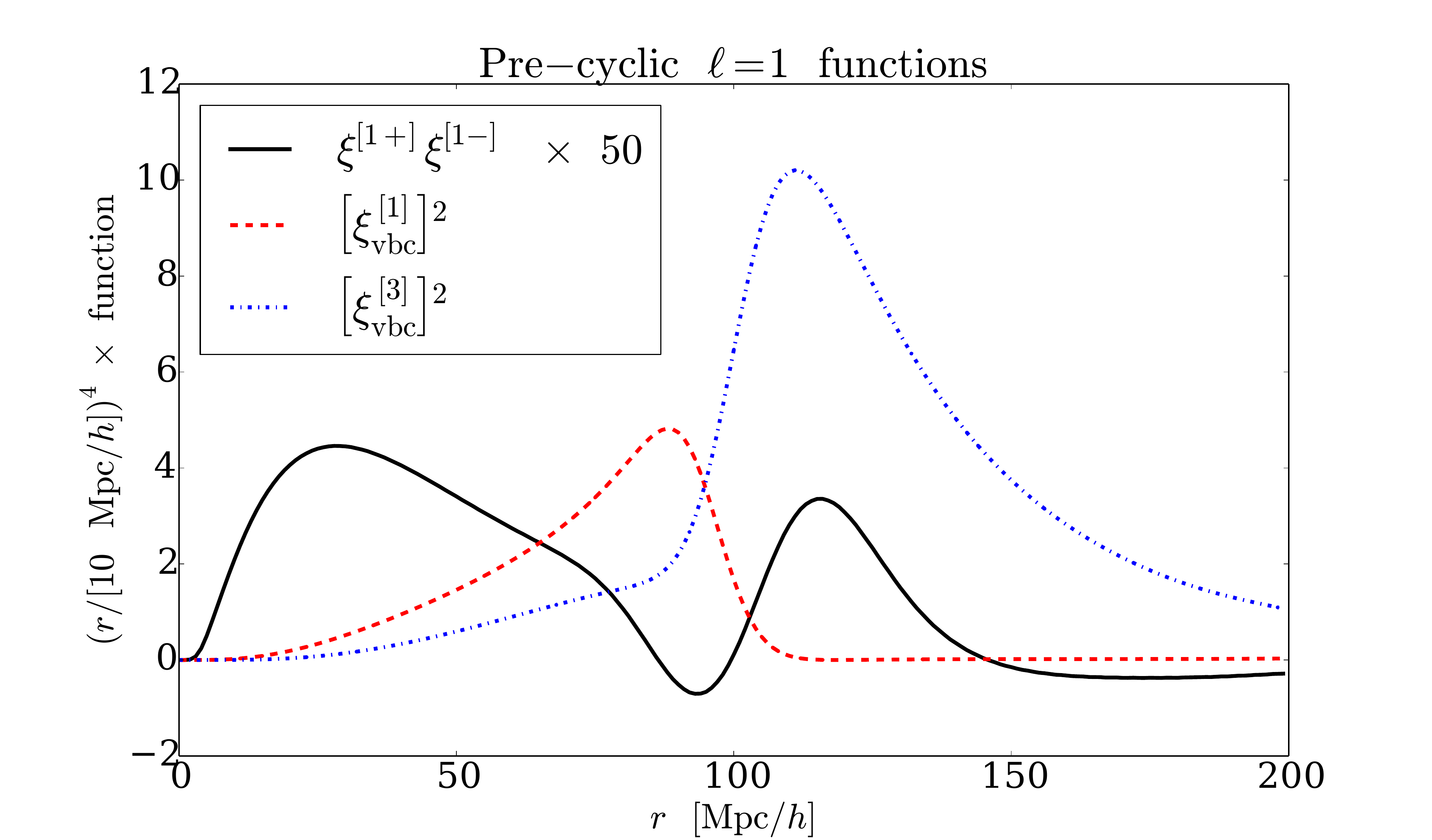}
\caption{The functions entering the pre-cyclic 3PCF at $\ell =1$ as in equation (\ref{eqn:config_space_model}), plotted for $r_1 = r_2 \equiv  r$. All functions are weighted by $r^4/[10\;\Mpch]^4$ to take out the large-scale fall-off of the 3PCF.  The first function, $\xi^{[1+]}\xi^{[1-]}$, has also been multiplied by fifty to be easily visible on the same scale as the other functions. All three functions have different structure around the BAO scale. In practice, since the $\ell=3$ function enters only at $\oO(\beta^2)$, it is not important for detecting the relative velocity. Consequently the truly essential point of this Figure is the difference in structure between the $\ell=1$ relative velocity contribution (dashed red) and the $\ell=1$ linear bias contribution (solid black). This  difference in structure suggests that the relative velocity will not be degenerate with the linear bias, as further discussed in SE15a.
}
\label{fig:pre_cyclic_ell_one_fns}
\end{figure}

\section{Conclusions \& Discussion}
\label{sec:concs}

In this work, we have shown how to transform the redshift-space bispectrum model of SCF99 into a configuration-space model suitable for fitting the 3PCF of current and upcoming large-scale structure redshift surveys.  Prior to cyclic summing, our configuration model space model can be written entirely in terms of simple 1-D transforms of the linear theory matter power spectrum.  We have also shown how to incorporate tidal tensor biasing in this framework, and developed a redshift-space model of the effect on the 3PCF of biasing by the baryon-dark matter relative velocity.  

Overall, a remarkably simple lesson emerges from this work: that the redshift-space 3PCF is to a fairly good approximation (of order $10\%$) simply a rescaling of the real-space 3PCF, where the rescaling factor is roughly independent of both triangle side length and multipole and depends on $\beta$ as well as the bias coefficients' values.  In principle, the model presented in this work should allow constraints to be placed not only on the bias values $b_1,\;b_2,\;b_t$, and $b_v$, but on $\beta$ itself and thence the matter density parameter $\Omega_{\rm m}$, since $\beta = f/b_1 \approx \Omega_{\rm m}^{0.55}/b_1$.  However, in practice $\beta$ is substantially degenerate with the linear bias.  Future work will explore what precision the 3PCF can offer on $\beta$ as well as on the other bias parameters. In this regard it may be worthwhile to compute higher moments of the 3PCF with respect to the line of sight, as these moments are $\oO(\beta)$ at lowest order and consequently may offer more sensitivity on $\beta$.  The mathematical techniques developed in SE15a and in this work are likely sufficient to transform the SCF99 expressions for the quadrupole moment of the 3PCF with respect to the line of sight; so doing is an avenue of possible future work.

An addtional avenue of future work would be joint fitting of the 2PCF multipoles and the 3PCF. In particular, the square of the 2PCF monopole scales as $\sigma_8^4 b_1^4[1+4/3\beta+\oO(\beta^2)]$, while at leading order in $\beta$ the 3PCF scales as $\sigma_8^4 b_1^3 [1+4/3\beta+\oO(\beta^2)]$, ignoring the non-linear bias, which enters with a slightly different $\beta$ dependence but empirically is found to be small for e.g. the CMASS sample in SDSS DR12 (Slepian et al. 2016a). Thus the ratio of the 2PCF monopole's square to the 3PCF reveals $b_1[1+\oO(\beta^2)]$. This ratio offers a good starting point for measuring $b_1$ using the 2PCF and 3PCF by largely eliminating the leading-order $\beta$-dependence.

This work also shows that the most significant BAO feature appears in the $\ell=1$ post-cyclic multipole of the 3PCF.  There is additional BAO information in the $\ell=2$ multipole, but this BAO information comes from the pre-cyclic $L=1$ multipole's contribution to the post-cyclic $\ell=2$. The multipoles $\ell > 3$ look rather similar to $\ell=3$, but $\ell = 0,\;1,\;2$ and $3$ all look distinctively different from each other.

There is also interest in removing a possible systematic shift in the BAO scale as measured from the 2PCF caused by galaxy biasing including a term in the baryon-dark matter relative velocity (Dalal, Pen \& Seljak 2010; Yoo, Dalal \& Seljak 2011; SE15a; Beutler et al. 2015; BMH16).  Here we have for the first time presented the relative velocity's contribution to the redshift-space 3PCF, and shown that it is likely non-degenerate with the other terms entering at $\ell=1$. This work should enable the 3PCF's use in constraining the relative velocity bias and removing any systematic shift it may induce of the BAO bump in the 2PCF.

We have already briefly discussed the treatments of RSD in other works modeling the 3PCF or bispectrum.  As regards the purely analytic, SCF99 and Rampf \& Wong (2012) are the current state of the art; fortunately though they use different flavors of perturbation theory (SPT vs. 2LPT) as Rampf \& Wong show they agree at leading order.  Rampf \& Wong (2012) also compute the one-loop correction to the redshift-space bispectrum (their equation (4.16)); it involves an exponential of the sum of Legendre polynomials of the cosine between the wave-vectors and the line of sight.  We leave for future work consideration of whether the mathematical techniques developed here can be used to convert this expression into a configuration-space model in terms of simple 1-D integral transforms of the power spectrum.

Recent observational works on the large-scale 3PCF have adopted different treatments of RSD, discussed in more detail in S16a \S5.2. S16a used the largest sample to date for a 3PCF measurement ($\sim 800,000$ LRGs in the SDSS DR12 CMASS sample), adopting a real-space bias model. The biases measured within this model were interpreted as effective quantitites reflecting the rescalinS16g RSD induce as well as the intrinsic biasing.  Gazta\~naga et al. (2009), the only other observational work to access the BAO scale, measured the 3PCF divided by the Peebles \& Groth (1977) hierarchical ansatz, and noted that on large scales RSD cancel out of the 3PCF. Consequently they did not incorporate any additional modeling of RSD. 

Gil-Mar\'in et al. (2015) measured the bispectrum of a sample similar in size to that of S16a ($\sim 600,000$ LRGs in the SDSS DR11 sample).  The starting point of that work was the bispectrum model of SCF99, but as further discussed in \S\ref{subsec:further_model}, they added additional parameters to reflect non-linear structure formation. These parameters are calibrated empirically from N-body simulations.  Gil-Mar\'in et al. (2015) also incorporated tidal tensor biasing but assumed the tidal tensor bias follows the theoretical relation for local Lagrangian biasing and consequently is fully determined by $b_1$; thus they did not independently fit for $b_t$.

Overall, there has not yet been a simple configuration-space model for the redshift-space 3PCF. The present work fills the gap, and we believe the template presented here will be of considerable utility in extracting BAO information from the 3PCF in current and future redshift surveys. Furthermore, we have developed a model for the relative velocity's contribution to the redshift-space 3PCF that allows a constraint on the relative velocity bias, important for ensuring the 2PCF remains an accurate avenue for measuring the cosmic expansion history. 

In a companion work (Slepian et al. 2016b) to this paper, we apply the templates presented here to the SDSS DR12 CMASS sample already used for the 3PCF measurement of S16a. We make a highly precise measurement of the linear bias as well as place constraints on the other bias parameters.  Furthermore, we make the first high-significance detection of the BAO in the 3PCF and use it to measure the cosmic distance scale to redshift $z=0.57$ with $1.7\%$ precision.  In a second companion paper (Slepian et al. 2016c), we use the relative velocity template presented here to constrain the relative velocity bias with $1\%$ precision, sufficient to imply that the BAO scale as measured from the SDSS DR12 2PCF is not systematically shifted. Future work might explore the utility of these 3PCF templates for even larger spectroscopic datasets, such as DESI (Levi et al. 2013).

\section*{Acknowledgements}
We thank Aaron Bray, Charles-Antoine Collins-Fekete, Tansu Daylan, Margaret Geller, Hector Gil-Mar\'in, Nick Hand, Nuala McCullagh, Stephen Portillo, Cornelius Rampf, Marcel Schmittfull, Uro\v{s} Seljak, Roman Scoccimarro, Joshua Suresh, Zvonimir Vlah, Martin White, and Alexander Wiegand for useful discussions. This material is based upon work supported by the National Science Foundation Graduate Research Fellowship under Grant No. DGE-1144152; DJE is supported by grant DE-SC0013718 from the U.S. Department of Energy.

\section*{Appendix A}
We wish to compute the inverse FT 
\begin{align}
I=&\int\frac{d^3\vk_1d^3\vk_2}{(2\pi)^6}\frac{1}{|\vk_1+\vk_2|^2}\sum_{\mathcal{L}} \mathcal{U}_{\mathcal L}(k_1,k_2)P_{\mathcal L}(x)\nonumber\\
&\times e^{-i\vk_1\cdot\vr_1}e^{-i\vk_2\cdot\vr_2}, 
\end{align}
where the kernel $\mathcal{U}$ is separable at each multipole as $\mathcal{U}_{\mathcal L}(k_1,k_2) = u_{\mathcal L}(k_1)u_{\mathcal L}(k_2)$. We can rewrite $I$ by introducing an additional degree of freedom $\vk_3$ and enforcing the constraint that $k_3^2=k_1^2+k_2^2+2k_1k_2x$ via an integral over $d^3\vk_3$ against a Dirac delta function, as
\begin{align}
I=&\int\frac{d^3\vk_1d^3\vk_2d^3\vk_3}{(2\pi)^9}\frac{1}{k_3^2}\sum_{\mathcal{L}}\mathcal{U}_{\mathcal L}(k_1,k_2)P_{\mathcal L}(x)\nonumber\\
&\times e^{-i\vk_1\cdot\vr_1}e^{-i\vk_2\cdot\vr_2}(2\pi)^3\delta_{\rm D}^{[3]}(\vk_1+\vk_2-\vk_3).
\end{align}
We now wish to factorize $I$ into a product of three integrals, each dependent on only one wavevector.  Using SE15b equation (58) to expand the Dirac delta function into spherical Bessel functions and spherical harmonics and also expanding the Legendre polynomial into spherical harmonics using the spherical harmonic addition theorem, we have
\begin{align}
I=&\sum_{\mathcal L}\sum_{l_1l_2l_3}\sum_{m_1m_2m_3}\mathcal{D}_{l_1l_2l_3}\mathcal{C}_{l_1l_2l_3}\left(\begin{array}{ccc}
l_{1} & l_{2} & l_{3}\\
0 & 0 & 0
\end{array}\right)\nonumber\\
&\times \left(\begin{array}{ccc}
l_{1} & l_{2} & l_{3}\\
m_{1} & m_{2} & m_{3}
\end{array}\right)\frac{4\pi}{2\mathcal{L}+1}\sum_{\mathcal{M}=-{\mathcal L}}^{\mathcal L}\int r^2 dr\nonumber\\
&\times\int\frac{d^3\vk_1}{2\pi^2}j_{l_1}(k_1r)u_{\mathcal L}(k_1)e^{-i\vk_1\cdot\vr_1}Y^*_{l_1m_1}(\hk_1)Y_{\mathcal L M}(\hk_1)\nonumber\\
&\times\int\frac{d^3\vk_2}{2\pi^2}j_{l_2}(k_2r)u_{\mathcal L}(k_2)e^{-i\vk_2\cdot\vr_2}Y^*_{l_2m_2}(\hk_2)Y^*_{\mathcal L M}(\hk_2)\nonumber\\
&\times \int\frac{d^3\vk_3}{2\pi^2}j_{l_3}(k_3r)\frac{1}{k_3^2}Y^*_{l_3m_3}(-\hk_3).
\label{eqn:I_intermediate}
\end{align}
We have defined $\mathcal{D}_{l_1 l_2 l_3} = i^{l_1+l_2+l_3}$ and $\mathcal{C}_{l_1 l_2 l_3}$ is defined below equation (\ref{eqn:interm_step2}). The integral over $d\Omega_3$ can be performed immediately by orthogonality, setting $l_3=0=m_3$ and reducing the integral over $d^3\vk_3$ to
\begin{align}
\sqrt{4\pi}\int\frac{dk_3}{2\pi^2}j_0(k_3r)=\frac{1}{\sqrt{4\pi}r}.
\end{align}
Had there been no factor of $1/k_3^2$, the $d^3\vk_3$ integral would have yielded $\sqrt{4\pi}\delta_{\rm D}^{[1]}(r)/(4\pi r^2)$, setting $r=0$ inside the $dr$ integral. In this case the $k_1$ and $k_2$ integrals would then be non-zero only for $l_1=0=l_2$ (recall that $j_n(0)=1$ for $n=0$ and zero otherwise). This case thus recovers the correct inverse FT were there only $k_1$ and $k_2$ dependence.

The 3j-symbols of equation (\ref{eqn:I_intermediate}) imply $l_1=l_2$ and $m_1 = -m_2$ and using NIST DLMF 34.3.1 can be evaluated as $(-1)^{m_1}/(2l_1+1)$.  The integrals over $d^3\vk_1$ and $d^3\vk_2$ can be split into angular and radial pieces by expanding the plane wave into spherical Bessel functions and spherical harmonics with indices $L_1M_1$ for that in $\vk_1\cdot\vr_1$ and $L_2M_2$ for that in $\vk_2\cdot\vr_2$. The angular pieces can then be evaluated in terms of 3j-symbols.  Manipulating the resulting 3j-symbols using NIST DLMF 34.3.10, summing them over $m_1$ and $\mathcal{M}$ using the orthogonality identity NIST DLMF 34.3.16 (which sets $L_2=L_1$ and $M_2=M_1$), invoking the spherical harmonic addition theorem, and simplifying yields
\begin{align}
&I_{\mathcal L}(r_1,r_2;\hr_1\cdot\hr_2)=\sum_{l_1 L_1}(-1)^{l_1+L_1}(2l_1+1)(2L_1+1)\nonumber\\
&\times \left(\begin{array}{ccc}
l_{1} & L_{1} & {\mathcal L}\\
0 & 0 & 0
\end{array}\right)^2 P_{L_1}(\hr_1\cdot\hr_2)\int r dr f^{\mathcal L}_{L_1 l_1}(r_1;r)f^{\mathcal L}_{L_1 l_1}(r_2;r)
\end{align}
where $f^{\mathcal L}_{L_1 l_1}(r_i;r)$ is defined as
\begin{align}
f^{\mathcal L}_{L_1 l_1}(r_i;r)=\int \frac{k^2dk}{2\pi^2}j_{L_1}(kr_i)j_{l_1}(kr)u_{\mathcal L}(k)P(k).
\label{eqn:ftens_gen_defn}
\end{align}
Finally, using orthogonality to project onto the $\ell^{th}$ multipole, we obtain 
\begin{align}
I_{{\mathcal L}\ell}(r_1,r_2) = &\sum_{l_1}(-1)^{l_1+\ell}(2l_1+1)(2\ell+1)\left(\begin{array}{ccc}
l_{1} & \ell & {\mathcal L}\\
0 & 0 & 0
\end{array}\right)^2\nonumber\\
&\times \int r dr f^{\mathcal L}_{\ell l_1}(r_1;r)f^{\mathcal L}_{\ell l_1}(r_2;r).
\label{eqn:IL_gen_result}
\end{align}

\section*{Appendix B}
We prove the identity that
\begin{align}
&j_L(k|\vr_1-\vr_2|)Y_{LM}(\reallywidehat{\vr_1-\vr_2})=4\pi\sum_{L_1M_1}\sum_{L_2M_2}i^{L_2-L_1+L}\nonumber\\
&\times j_{L_1}(kr_1)j_{L_2}(kr_2)\mathcal{C}_{L_1 L_2 L}\left(\begin{array}{ccc}
L_{1} & L_{2} & L\\
0 & 0 & 0
\end{array}\right)\nonumber\\
&\times \left(\begin{array}{ccc}
L_{1} & L_{2} & L\\
M_1 & M_2 & M
\end{array}\right) Y_{L_1 M_1}^*(\hr_1)Y_{L_2 M_2}^*(\hr_2),
\label{eqn:sep_id}
\end{align}
where $\mathcal{C}_{L_1 L_2 L}$ is defined below equation (\ref{eqn:interm_step2}).
We begin with the equality
\begin{align}
e^{-i\vk\cdot(\vr_1-\vr_2)}=e^{-i\vk\cdot\vr_1}e^{i\vk\cdot\vr_2}
\end{align}
and apply the plane wave expansion once on the lefthand side and twice on the righthand side to find
\begin{align}
&4\pi \sum_{L'M'}(-i)^{L'}j_{L'}(k|\vr_1-\vr_2|)Y_{L' M'}^*(\hk)Y_{L' M'}(\reallywidehat{\vr_1-\vr_2})\nonumber\\
&=(4\pi)^2\sum_{L_1 M_1}\sum_{L_2 M_2} i^{L_2-L_1}j_{L_1}(kr_1)j_{L_2}(kr_2)\nonumber\\&\times Y_{L_1 M_1}(\hk)Y_{L_2 M_2}(\hk)Y^*_{L_1 M_1}(\hr_1)Y^*_{L_2 M_2}(\hr_2).
\end{align}
We now integrate both sides against $Y_{LM}(\hk)$, invoking orthogonality on the lefthand side and using 3j-symbols on the righthand side.  Rearranging what results yields the desired equation (\ref{eqn:sep_id}); it is equivalent to Mehrem (2002) equation (4.14) if his Clebsch-Gordan symbols are translated to 3j-symbols and $\vr_2\to -\vr_2$.

\section*{References}

\hangindent=1.5em
\hangafter=1
\noindent Baldauf T., Seljak U., Desjacques V. \& McDonald P., 2012, PRD 86, 8.

\hangindent=1.5em 
\hangafter=1
\noindent Bernardeau F., Colombi S., Gazta\~{n}aga E., Scoccimarro R., 2002, Phys. Rep., 367, 1.

\hangindent=1.5em 
\hangafter=1
\noindent Blake C. \& Glazebrook K., 2003, ApJ 594, 2, 665-673.

\hangindent=1.5em
\hangafter=1
\noindent Chan K.C., Scoccimarro R. \& Sheth R.K., 2012, PRD 85, 8, 083509.

\hangindent=1.5em
\hangafter=1
\noindent Cuesta A.J. et al., 2015, MNRAS 457, 2, 1770-1785. 

\hangindent=1.5em
\hangafter=1
\noindent Dawson K.S. et al., 2013, AJ 145, 1, 10.

\hangindent=1.5em
\hangafter=1
\noindent Drinkwater M.J. et al., 2010, MNRAS 401, 3, 1429-1452. 

\hangindent=1.5em
\hangafter=1
\noindent Eisenstein D.J. \& Hu W., 1998, ApJ 496, 605.

\hangindent=1.5em
\hangafter=1
\noindent Eisenstein D.J., Hu W. \& Tegmark M., 1998, ApJ  504:L57-L60.

\hangindent=1.5em
\hangafter=1
\noindent Eisenstein D.J., Seo H.-J. \& White M., 2007, ApJ 664, 2, 660-674.

\hangindent=1.5em
\hangafter=1
\noindent Eisenstein D.J. et al., 2011, AJ 142, 3, 72.

\hangindent=1.5em
\hangafter=1
\noindent Gil-Mar\'in H., Nore\~na J., Verde L., Percival W.J., Wagner C., Manera M. \& Schneider D.P., 2015, MNRAS 451, 1, 539-580.

\hangindent=1.5em
\hangafter=1
\noindent Gil-Mar\'in H., Wagner C., Nore\~na J., Verde L. \& Percival W., 2014, JCAP 12, 029.

\hangindent=1.5em
\hangafter=1
\noindent Gil-Mar\'in H.,  Wagner C., Verde L., Percival W.J., Wagner C., Manera M. \& Schneider D.P., 2015, MNRAS 451, 1, 539-580.

\hangindent=1.5em
\hangafter=1
\noindent Groth E. J. \& Peebles P. J. E., 1977, 217, 385.

\hangindent=1.5em 
\hangafter=1
\noindent Hamilton A.J.S., 1992,  ApJL 385, L5-L8.

\hangindent=1.5em 
\hangafter=1
\noindent Hamilton A.J.S, 1998, in ``The Evolving Universe: Selected Topics on Large-Scale Structure and on the Properties of Galaxies,'' Dordrecht: Kluwer.

\hangindent=1.5em 
\hangafter=1
\noindent Hivon E., Bouchet F.R., Colombi S. \& Juszkiewicz R., 1995, A\&A 298, 643.

\hangindent=1.5em 
\hangafter=1
\noindent Hu W. \& Haiman Z., 2003, PRD 68, 6, 063004.

\hangindent=1.5em 
\hangafter=1
\noindent Jackson J.C., 1972, MNRAS 156, 1.

\hangindent=1.5em
\hangafter=1
\noindent Jones D.H. et al., 2009, MNRAS  399, 2, 683-698.

\hangindent=1.5em
\hangafter=1
\noindent Kaiser N., 1987, MNRAS 227, 1.

\hangindent=1.5em
\hangafter=1
\noindent Lewis A., 2000, ApJ, 538, 473.

\hangindent=1.5em 
\hangafter=1
\noindent Linder E.V., 2003, PRD 68, 8, 083504.

\hangindent=1.5em 
\hangafter=1
\noindent Ma, C.-P. \& Fry, J. N. 2000, ApJ, 531, L87-L90.

\hangindent=1.5em
\hangafter=1
\noindent McDonald P. \& Roy A., 2009, JCAP 0908, 020.

\hangindent=1.5em
\hangafter=1
\noindent Mehrem R., 2002, preprint (arXiv:0909.0494v4). 

\hangindent=1.5em 
\hangafter=1
\noindent Olver W. J., Lozier D. W., Boisvert R. F., Clark C. W. eds. NIST Handbook
of Mathematical Functions. Cambridge University Press, Cambridge,\\
Available at http://dlmf.nist.gov/

\hangindent=1.5em
\hangafter=1
\noindent Scoccimarro R., Couchman H.M.P. \& Frieman J.A., 1999, ApJ 517:531-540.

\hangindent=1.5em
\hangafter=1
\noindent Sefusatti E., Crocce M., Pueblas S. \& Scoccimarro R., 2006, PRD 74, 2, 023522

\noindent Seo H.J. \& Eisenstein D.J., 2003. ApJ 598, 2, 720-740.

\hangindent=1em
\hangafter=1
\noindent Slepian Z. \& Eisenstein D.J., 2015a, MNRAS 448, 1, 9-26.

\hangindent=1em
\hangafter=1
\noindent Slepian Z. \& Eisenstein D.J., 2015b, MNRAS 454, 4, 4142-4158.

\hangindent=1em
\hangafter=1
\noindent Slepian Z. \& Eisenstein D.J., 2015c, MNRASL 455, 1, L31-L35.

\hangindent=1em
\hangafter=1
\noindent Slepian Z. \& Eisenstein D.J., 2016, MNRAS 457, 24-37. 

\hangindent=1em
\hangafter=1
\noindent Slepian Z.  et al., 2016a, preprint (arXiv:1512.02231). 

\hangindent=1em
\hangafter=1
\noindent Slepian Z.  et al., 2016b, preprint (arXiv:---). 

\hangindent=1em
\hangafter=1
\noindent Slepian Z.  et al., 2016c, preprint (arXiv:---). 

\hangindent=1em
\hangafter=1
\noindent Smith R.E., Sheth R.K. \& Scoccimarro R., 2008, PRD 78, 2, 023523.

\hangindent=1em 
\hangafter=1
\noindent Verde L., Heavens A.F. \& Matarrese S., 1998, MNRAS 300, 3, 747-756.

\end{document}